\documentclass{aa}  

\usepackage{graphicx}
\usepackage{txfonts}
\usepackage{color}

\begin{document}

   \title{CO envelope of the symbiotic star R Aqr seen by ALMA}

   \author{S. Ramstedt
          \inst{1}
          \and
          S.~Mohamed\inst{2,3,4} \and T.~Olander\inst{1} \and W.~H.~T. Vlemmings\inst{5} \and T.~Khouri\inst{5} \and S.~Liljegren\inst{1}
          }

   \institute{Division for Astronomy and Space Physics, Department for Physics and Astronomy, Uppsala University, Box 512, 753 20 Uppsala, Sweden\\
              \email{sofia.ramstedt@physics.uu.se} \and
              South African Astronomical Observatory, P.~O.~Box 9, 7935 Observatory, South Africa \and
              Astronomy Department, University of Cape Town, 7701 Rondebosch, South Africa \and
              South African National Institute for Theoretical Physics, Private Bag X1, 7602 Matieland, South Africa \and
              Department of Space, Earth and Environment, Chalmers University of Technology, Onsala Space Observatory, 439 92 Onsala, Sweden
             }

   \date{Received ; accepted }

%context
%aims
%methods
%results
%conclusions

\abstract{The symbiotic star R~Aqr is part of a small sample of binary AGB stars observed with the Atacama Large Millimeter/submillimeter Array (ALMA). The sample stars are: R~Aqr, Mira, W~Aql, and $\pi^{1}$~Gru. The sample covers a range in binary separation and wind properties, where R~Aqr is the source with the smallest separation. The R~Aqr binary pair consists of an M-type AGB star and a white dwarf at a separation of 45\,mas, equivalent to about 10\,AU at 218\,pc. The aim of the ALMA study is to investigate the dependence of the wind shaping on the binary separation and to provide constraints for hydrodynamical binary interaction models. R~Aqr is particularly interesting as the source with the smallest separation and a complex circumstellar environment that is strongly affected by the interaction between the two stars and by the high-energy radiation resulting from this interaction and from the hot white dwarf companion.

The CO($J$\,=\,3$\rightarrow$2) line emission has been observed with ALMA at $\sim$0.5\arcsec spatial resolution. The CO envelope around the binary pair is marginally resolved, showing what appears to be a rather complex distribution. The outer radius of the CO emitting region is estimated from the data and found to be about a factor of 10 larger than previously thought. This implies an average mass-loss rate during the past $\sim$100\,yr of $\dot{M}\approx2\times$10$^{-7}$\,M$_{\odot}$\,yr$^{-1}$, a factor of 45 less than previous estimates. The channel maps are presented and the molecular gas distribution is discussed and set into the context of what was previously known about the system from multiwavelength observations. Additional molecular line emission detected within the bandwidth covered by the ALMA observations is also presented. 

Because of the limited extent of the emission, firm conclusions about the dynamical evolution of the system will have to wait for higher spatial resolution observations. However, the data presented here support the assumption that the mass-loss rate from the Mira star strongly varies and is focused on the orbital plane.}
 

   \maketitle
%
%-------------------------------------------------------------------

\section{Introduction}
Symbiotic stars are binary stars that are identified through their composite spectra. The binary pair has one hotter compact component, such as~a white dwarf, and a cooler, evolved red giant, for example,~an asymptotic giant branch (AGB) star. At 218\,pc \citep{minetal14}, R~Aqr is the closest symbiotic star, and it is therefore well studied across several wavelengths \citep[e.g.,][and references therein]{ragletal06,kelletal07,mayeetal13,melnetal18}. It is one of very few symbiotics in which the binary pair has been resolved \citep[e.g.,][]{schmetal17}, and for which constraints on the orbital parameters are available \citep[orbital period of $\sim$16000 days; semimajor axis of $\sim$15.5\,AU,][]{grommiko09}.  Symbiotic stars are the proposed progenitor systems of planetary nebulae (PNe) \citep{joneboff17} and possibly also of supernovae type-Ia \citep[e.g.,][]{hamuetal03}. As the closest symbiotic, R~Aqr is studied to determine the dynamical processes behind these important astrophysical phenomena. 

%The stellar wind during the AGB builds up an expanding circumstellar envelope (CSE) consisting mainly of molecular gas and dust. The wind expansion velocity is larger than the escape velocity from the stellar surface and the CSE is unbound, continues to expand, and later forms the PNe. Therefore, t
To understand the transition from the AGB to the PN phase, observations of the molecular gas are particularly important. R~Aqr is part of a small sample of binary AGB stars observed with ALMA to provide observational constraints for hydrodynamical models of the interaction between the wind of an AGB star and a companion \citep[e.g.,][]{mohapods12}. Results on the other sample sources, Mira, W~Aql, and $\pi^{1}$~Gru, have been published in \citet{ramsetal14}, \citet{ramsetal17}, \citet{doanetal17}, \citet{khouetal16}, \citet{debeetal17}, and \citet{brunetal18}. R~Aqr is the source with the smallest separation in the sample and is also the source whose CO distribution seems most strongly affected by the hard radiation in the system.

CO line emission toward R~Aqr was first detected by \citet{bujaetal10}, who observed the $^{12}$CO $J$\,=\,2\,$\rightarrow$\,1 and 1\,$\rightarrow$\,0 emission with the IRAM\,30m telescope. They modeled the spatially unresolved line emission and derived an average mass-loss rate of $\dot{M}$=9$\times$10$^{-6}$\,M$_{\odot}$\,yr$^{-1}$  assuming that the size of the CO envelope is on the order of the binary separation (2$\times$10$^{14}$\,cm). The model reproduces the shape of the $J$\,=\,2\,$\rightarrow$\,1 line, including self-absorption in the approaching gas.

In this paper we present the first resolved CO line observations of the circumstellar envelope (CSE) around R~Aqr. In Sect.~\ref{obs} the observational setup is presented. In Sect.~\ref{data} the data are analyzed and the results are shown and discussed. Finally, in Sect.~\ref{context} the molecular gas distribution is compared to what is previously known from observations of the atomic gas. We conclude in Sect.~\ref{conc}.
%--------------------------------------------------------------------

   \begin{figure*}
   \centering
   \includegraphics[height=5.45cm]{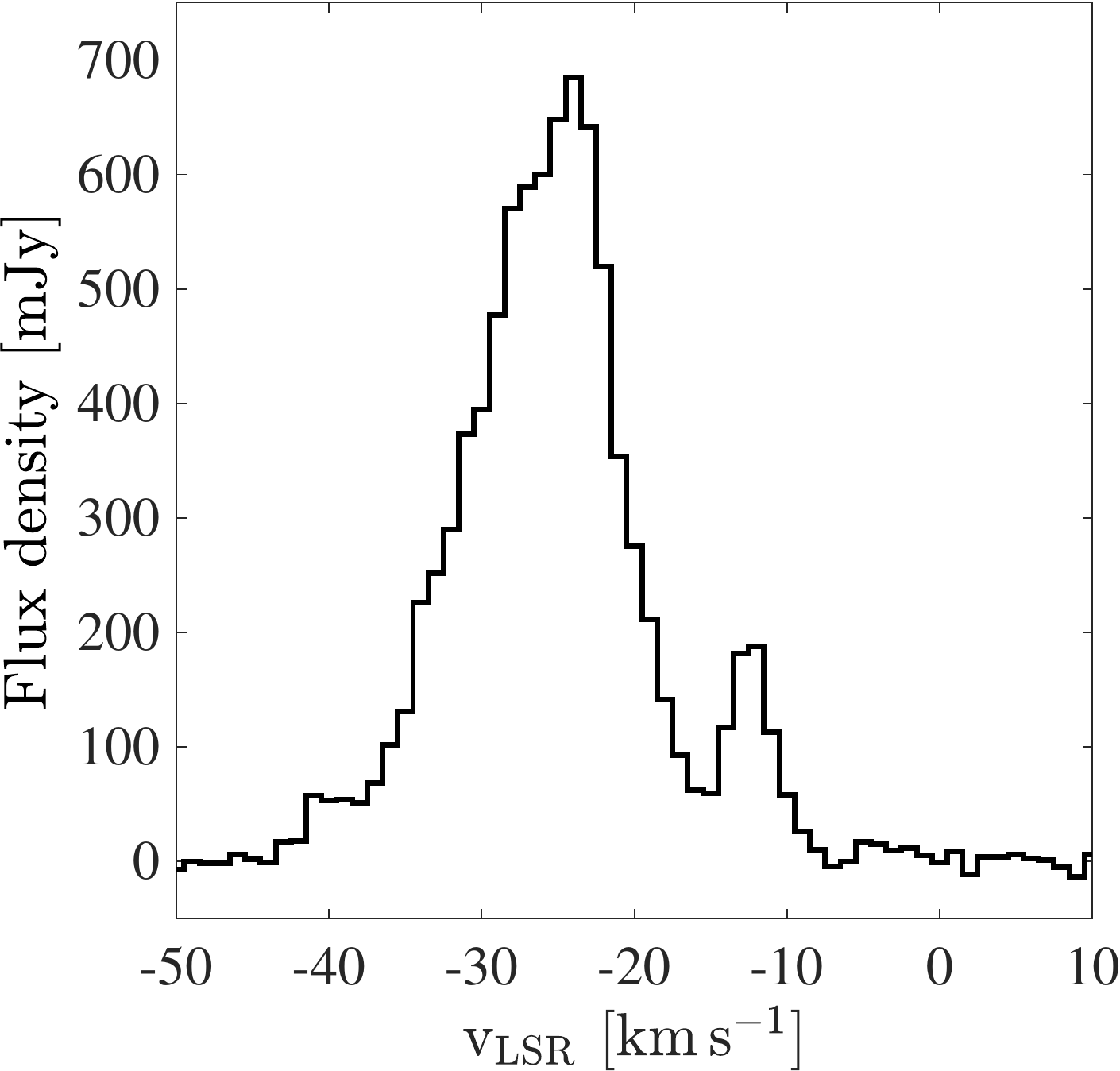}
   \includegraphics[height=5.5cm]{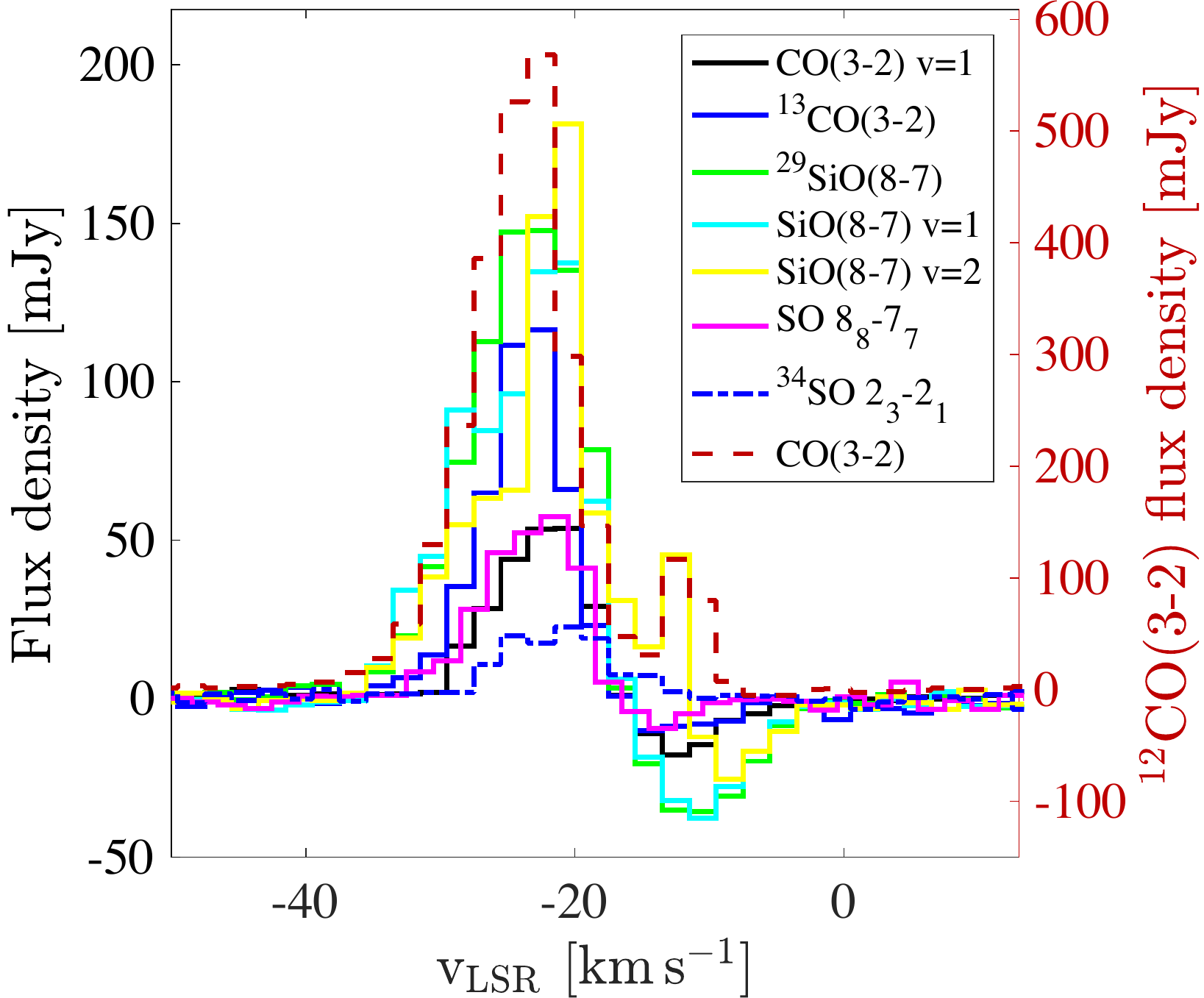}
   \includegraphics[height=5.7cm]{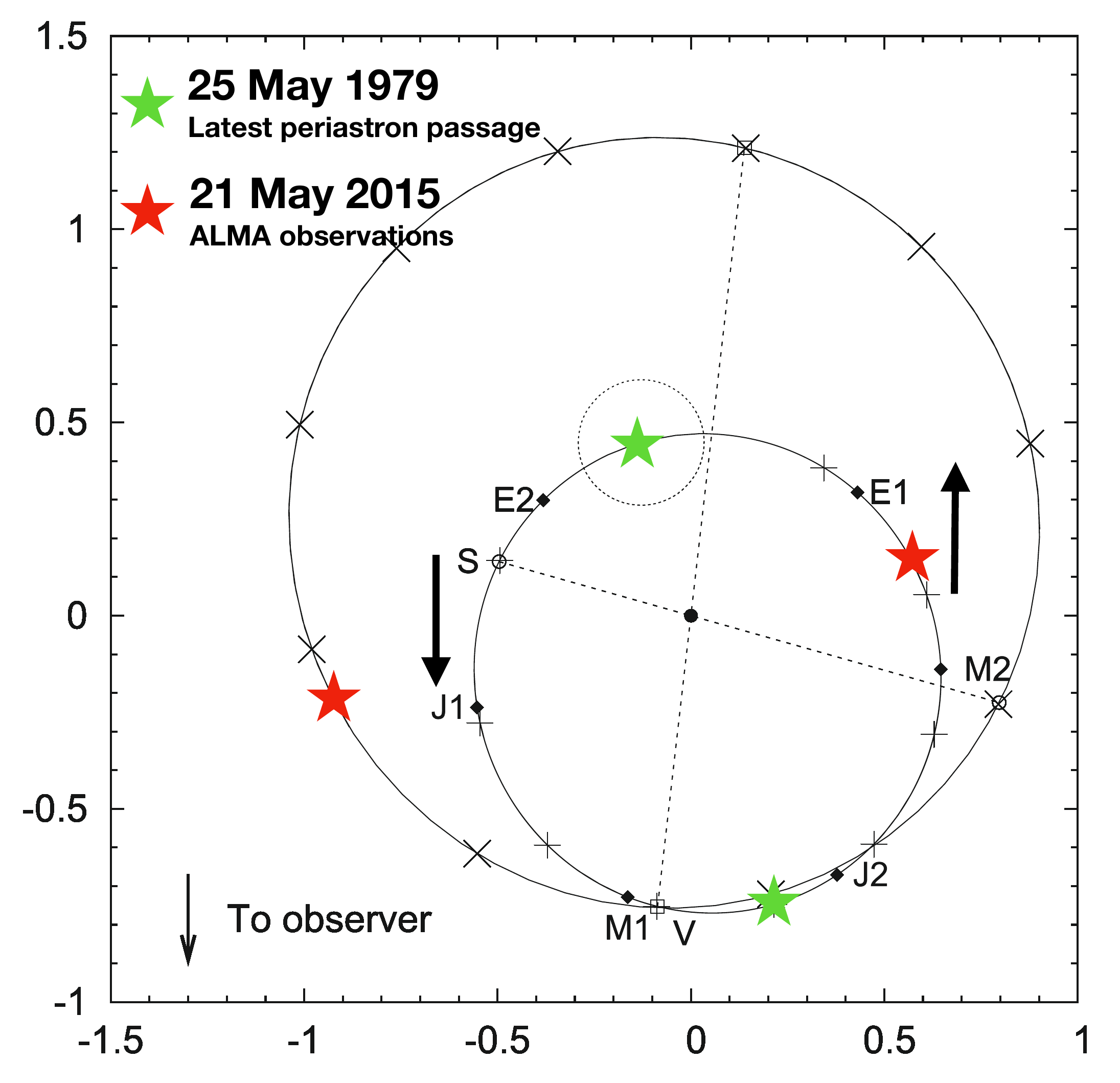}
   \caption{{\bf Left:}  $^{12}$CO(3-2) line detected toward R~Aqr with ALMA. The line is generated from channel maps with 1\,km\,s$^{-1}$ spectral resolution and measured across a 3\arcsec~aperture centered on the continuum emission peak. {\bf Middle:} Line emission from the lines listed in Table~\ref{lines}. All lines (including $^{12}$CO(3-2)) are generated at one pixel coincident with the continuum emission peak, and at 2\,km\,s$^{-1}$ spectral resolution. {\bf Right:} Orbits of the binary pair as derived by \citet{grommiko09}. Both the Mira star (+) and the hot component (x) move anticlockwise in this representation, and the center of mass is marked by the solid dot. The axes unit is the system semimajor axis, $a$ \citep[$\sim$15.5\,AU according to][]{grommiko09}. For details see \citet{grommiko09}. The position of the stars at the latest periastron passage (green stars) and the time of the ALMA observations (red stars) have been added. The base of the thick arrows marks the times of maximum velocity approach and retreat of the Mira star from the observer.}
   \label{COline}%
   \end{figure*}

\section{Observations}
\label{obs}
The R~Aqr system was observed on 21 May 2015 with 36 main-array 12\,m antennas. From previous observations \citep{bujaetal10}, the full extent of the CO envelope was expected to be smaller than the largest recoverable scale of the main array at the observed frequencies, and no mosaicing, Atacama Compact Array (ACA), or total power observations were performed. The longest baseline of the main array was 539\,m, and the shortest was 21\,m, corresponding to a maximum recoverable scale of just over 5\arcsec at 345\,GHz.

The correlator was set up with four $\sim$2\,GHz wide spectral windows centered on 331, 333, 343, and 345\,GHz. Both the $^{12}$CO and $^{13}$CO $J$\,=\,3\,$\rightarrow$\,2 lines were covered by this setup, together with several other chemically interesting circumstellar lines \citep[see][for the analysis of these lines from W~Aql]{brunetal18}. The spectral resolution of the original data was 0.488\,MHz.
 
The total on-source observing time was 48.15 mins. Standard calibration and imaging was performed using the Common Astronomy Software Applications package \citep[CASA,][]{mcmuetal07}. Ceres and quasar J2348-1631 were used as amplitude and bandpass calibrators. Phase calibration was carried out on quasar J2348-1631. A continuum image was made from the emission-free channels across all spectral windows. The position and flux at 338\,GHz was determined from image fitting and is given in Table~\ref{cont}. Table~\ref{cont} gives the formal fitting error, but additional flux uncertainties exist that are based on the accuracy of the solar system flux models\footnote{Butler~B., 2012, ALMA Memo 594}. The Ceres fluxes for the observing epoch were checked in CASA 5.1.1 and was found to be 13\% lower than adopted in the pipeline reduction. We therefore rescaled all our reported flux values (also the line fluxes). Based on an investigation of the fluxes derived for J2348-1631 compared with the monitoring in the ALMA calibrator catalog, we caution that the absolute flux calibration error using Ceres could still be on the order of 20\% . The continuum was subtracted from the channels covering the $^{12}$CO $J$\,=\,3\,$\rightarrow$\,2 (henceforth referred to as CO(3-2)) line emission before spectral averaging to 2\,km\,s$^{-1}$ , and final imaging was performed. The full width at half-maximum beam width of the final images is 0\farcs38$\times$0\farcs33 with a position angle of 77.4$^{\circ}$. The rms noise level reaches 4.0\,mJy/beam in the emission-free channels. Since this is the first detection of the CO(3-2) line emission from the system (to our knowledge),  no data exist that can be used to evaluate whether the total flux was recovered by the interferometer, but the extent of the CO emission appears much smaller than 5\arcsec~, as we discuss in more detail below. The CO(3-2) line emission peaks at a local standard of rest (LSR) velocity of $v_{\rm{LSR}}$=-24\,km\,s$^{-1}$ (Fig.~\ref{COline}).

\begin{table}[h]
\caption{Current position (J2000.0) and continuum flux density at 338\,GHz measured from the ALMA data. The flux error is the formal fitting error. See text for further discussion.}
\begin{center}
\begin{tabular}{lccc}
\hline \hline
Source & RA & Dec & $S_{\nu}$ \\
\hline
R~Aqr & 23:43:49.491 & -15:17:04.648 & 95.3$\pm$0.6\,mJy \\
\hline
\end{tabular}
\end{center}
\label{cont}
\end{table}%

Figure~\ref{COline} (left) shows the CO(3-2) line emission. The line is generated at 2\,km\,s$^{-1}$ spectral resolution and measured with an 3\arcsec~aperture centered on the peak of the continuum emission. The frequency, peak, and integrated flux density measured with a 3\arcsec~aperture of all the detected lines are listed in Table~\ref{lines}, and the line profiles other than $^{12}$CO(3-2) are shown in Fig.~\ref{COline} (middle). Only the $^{12}$CO(3-2) line emission is spatially resolved. All the other detected lines show unresolved emission.

\begin{table}[h]
\caption{Frequency, peak, and integrated flux density measured at one pixel coincident with the continuum emission peak of all lines at 2\,km\,s$^{-1}$ spectral resolution detected with ALMA within the observed spectral windows.}
\begin{center}
\begin{tabular}{lcccc}
\hline \hline
Transition && $\nu$ & $S^{\rm{peak}}_{\nu}$ & $S_{\nu}$ \\
 && [GHz] & [mJy] & [Jy\,km\,s$^{-1}$] \\
\hline
\hline
$^{12}$CO $J$=3-2, $v$=0 && 345.796 & 570 & 5.4\phantom{0} \\
$^{12}$CO $J$=3-2, $v$=1 && 342.648 & \phantom{1}54 & 0.30 \\
$^{13}$CO $J$=3-2, $v$=0 && 330.588 & 120 & 0.70 \\
$^{29}$SiO $J$=8-7, $v$=0 && 342.981 & 150 & 1.3\phantom{0}  \\
$^{28}$SiO $J$=8-7, $v$=1 && 342.504 & 140 &  1.1\phantom{0} \\
$^{28}$SiO $J$=8-7, $v$=2 && 344.916 & 180 &  1.3\phantom{0} \\
$^{32}$SO $N_J$=8$_{8}$-7$_{7}$, $v$=0 && 344.311 & \phantom{1}57 & 0.44 \\
$^{34}$SO $N_J$=2$_{3}$-2$_{1}$, $v$=0 && 343.851 & \phantom{1}23 & 0.23 \\
\hline
\hline
\end{tabular}
\end{center}
\label{lines}
\end{table}%

   \begin{figure*}
   \centering
   \includegraphics[width=4.75cm]{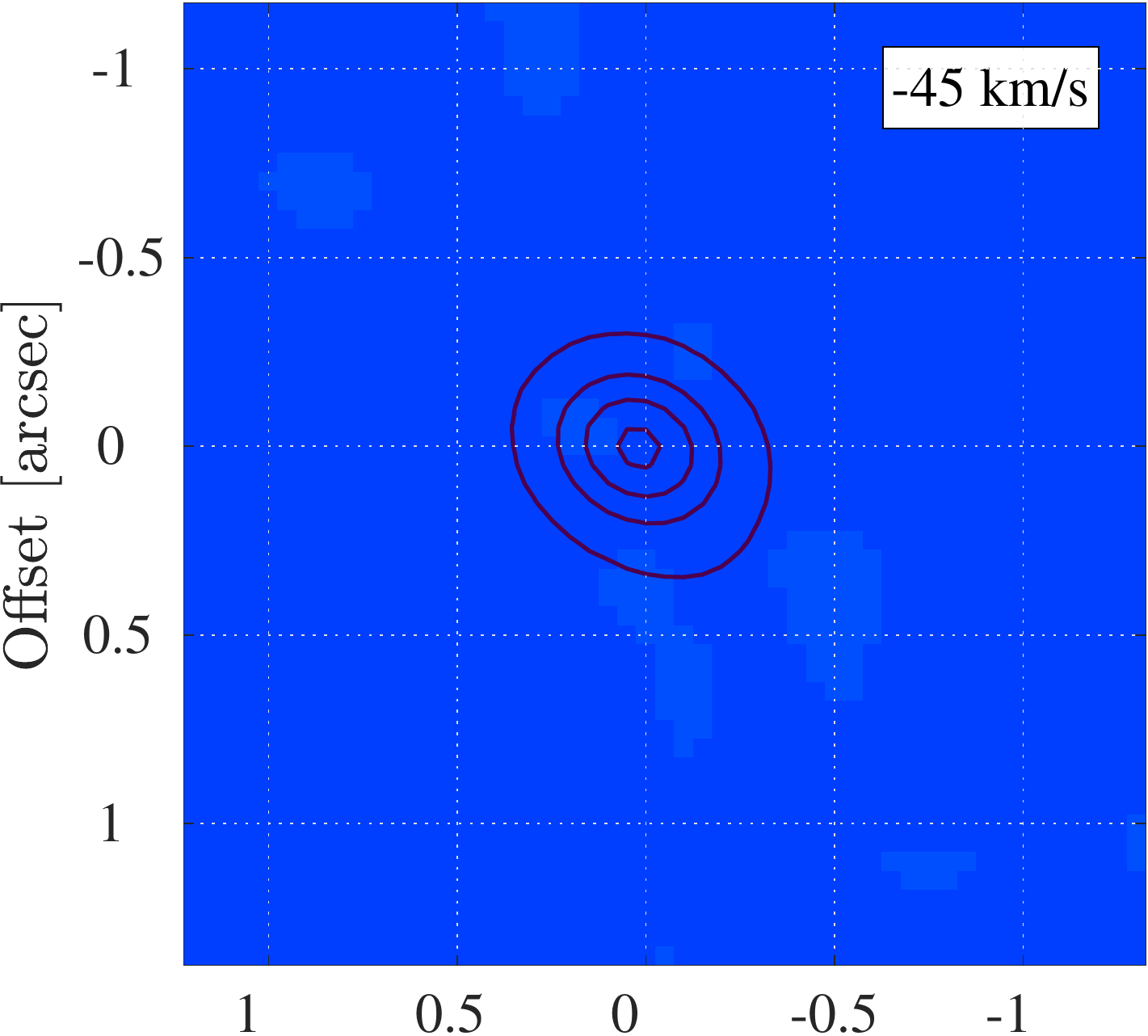}
   \includegraphics[width=4.43cm]{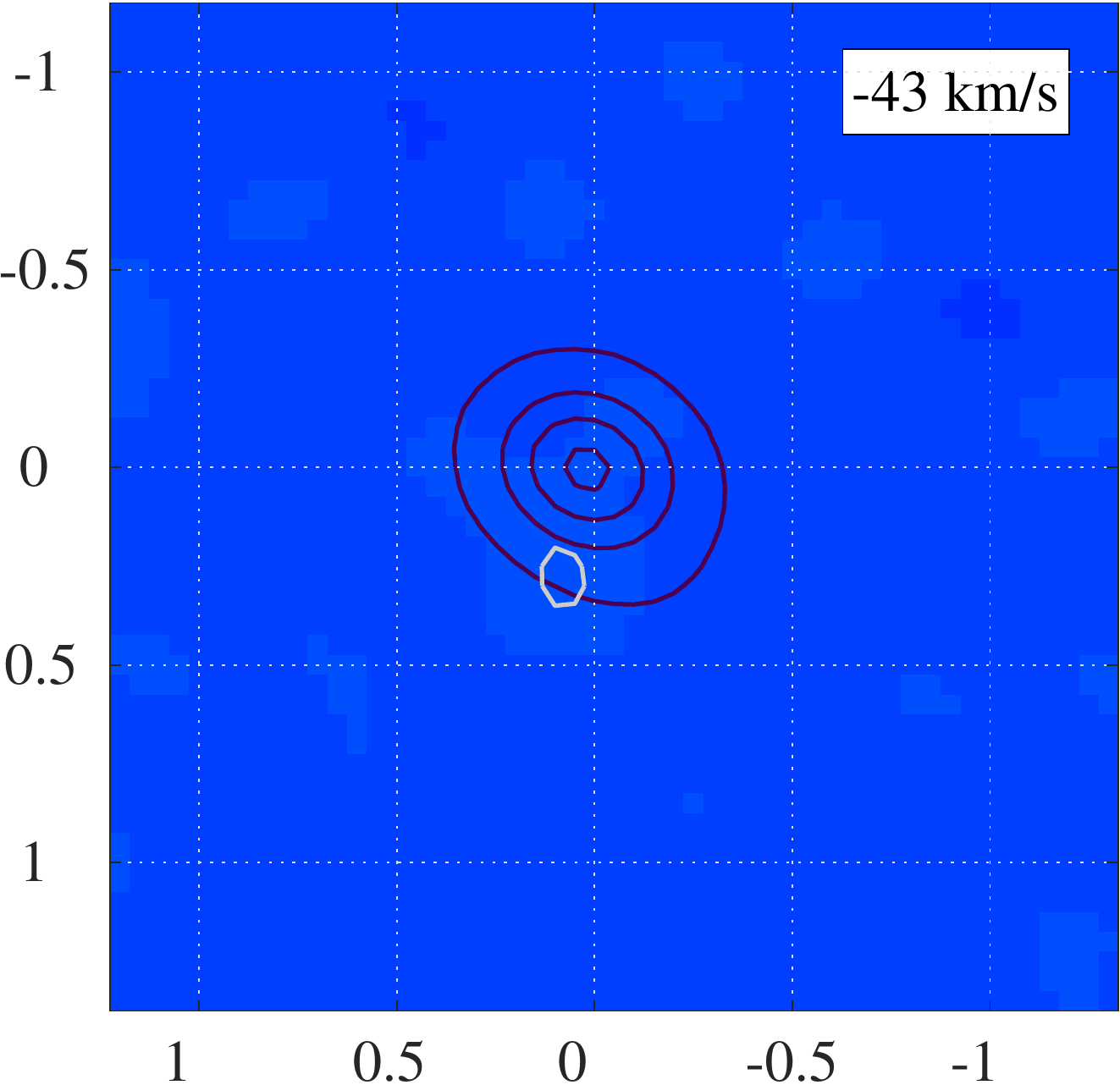}
   \includegraphics[width=4.43cm]{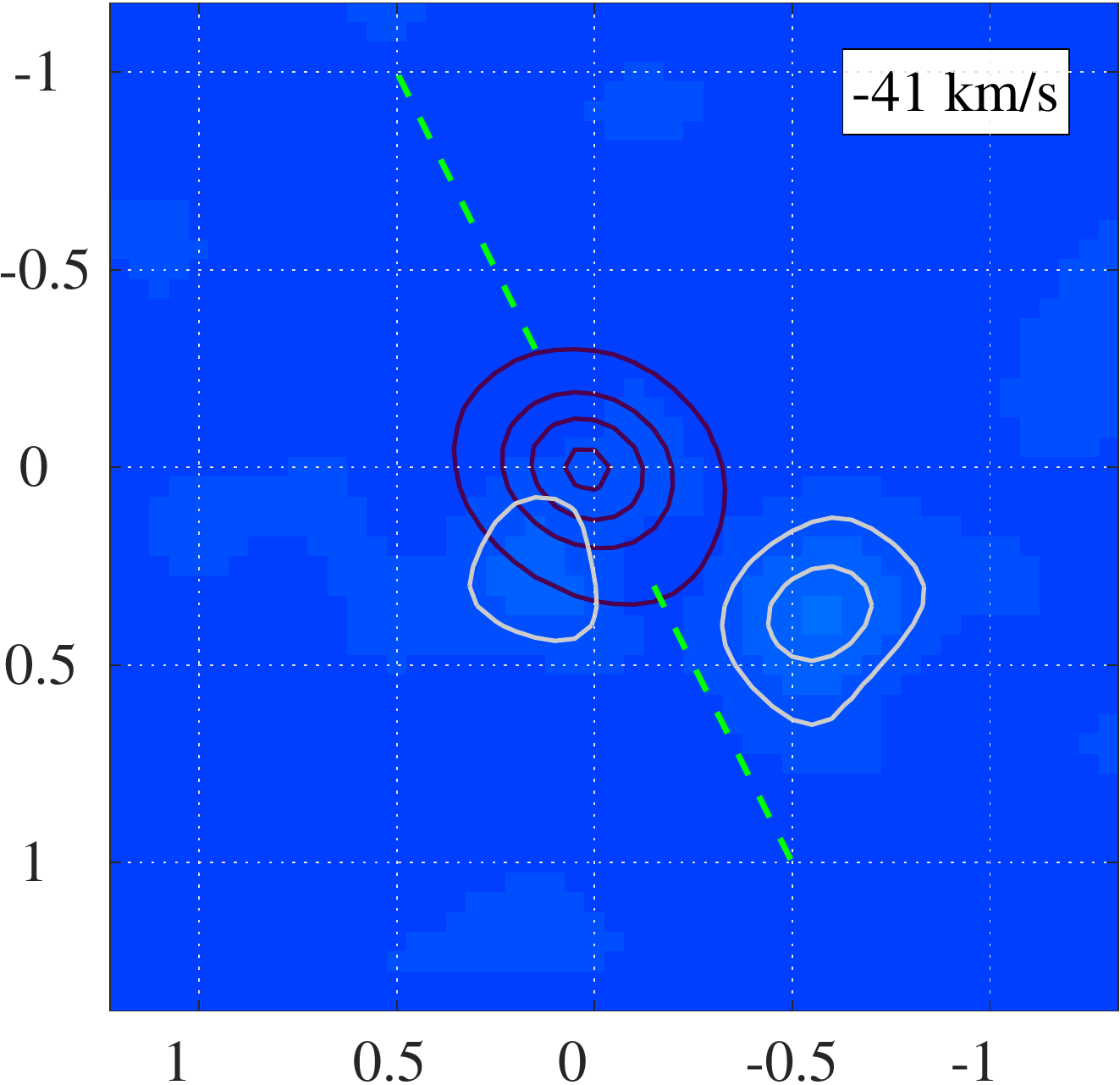}
   \includegraphics[width=4.43cm]{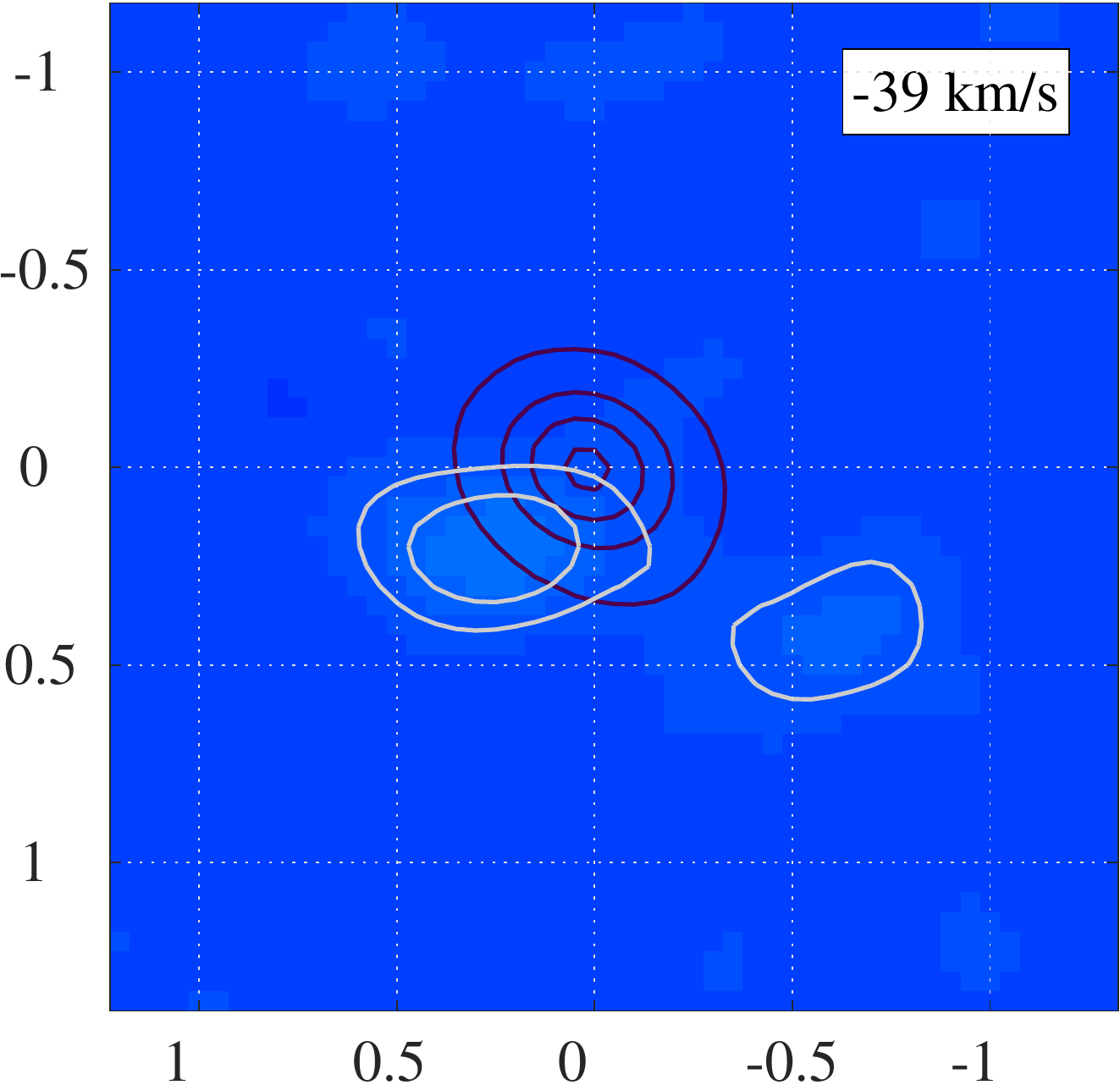}
   \includegraphics[width=4.75cm]{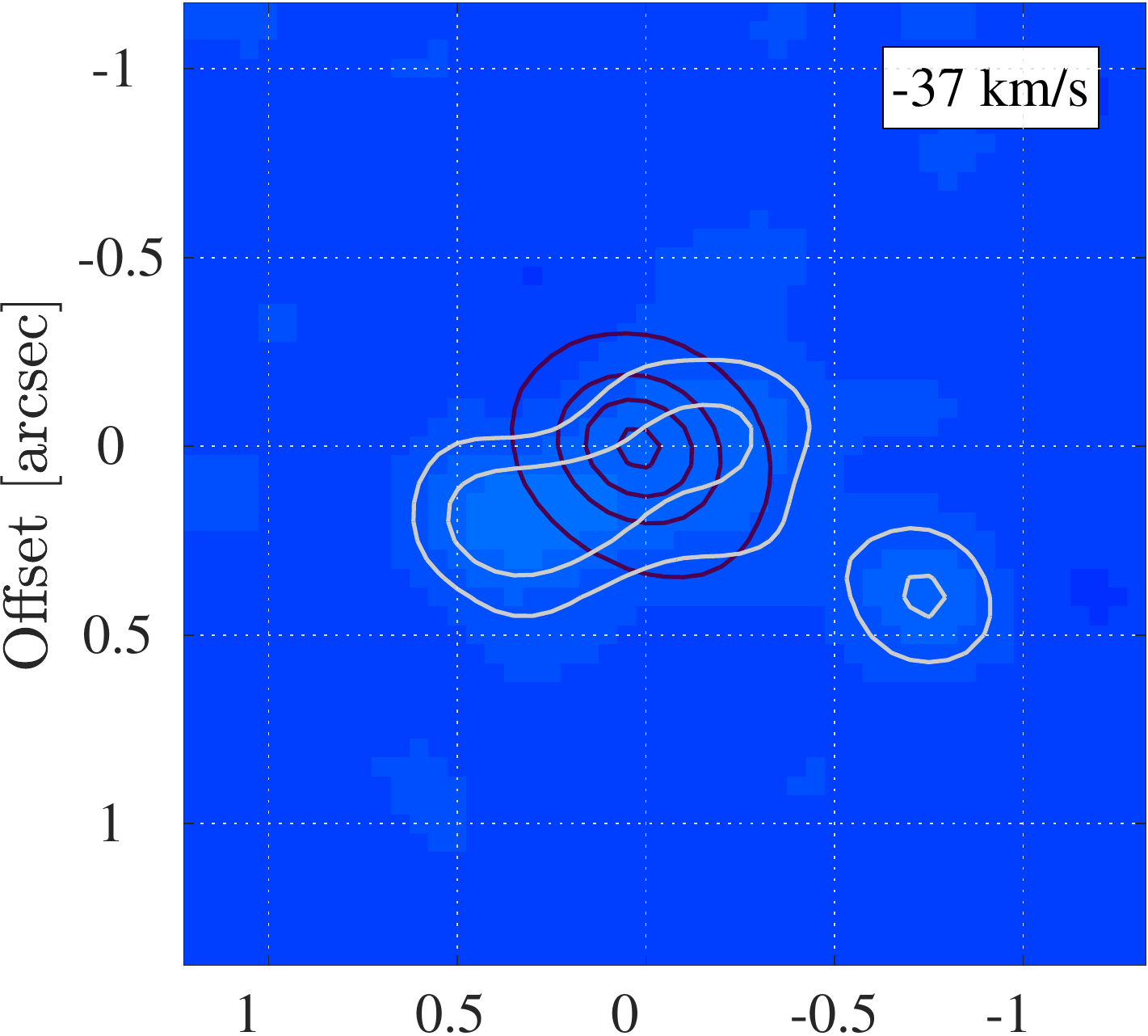}
   \includegraphics[width=4.43cm]{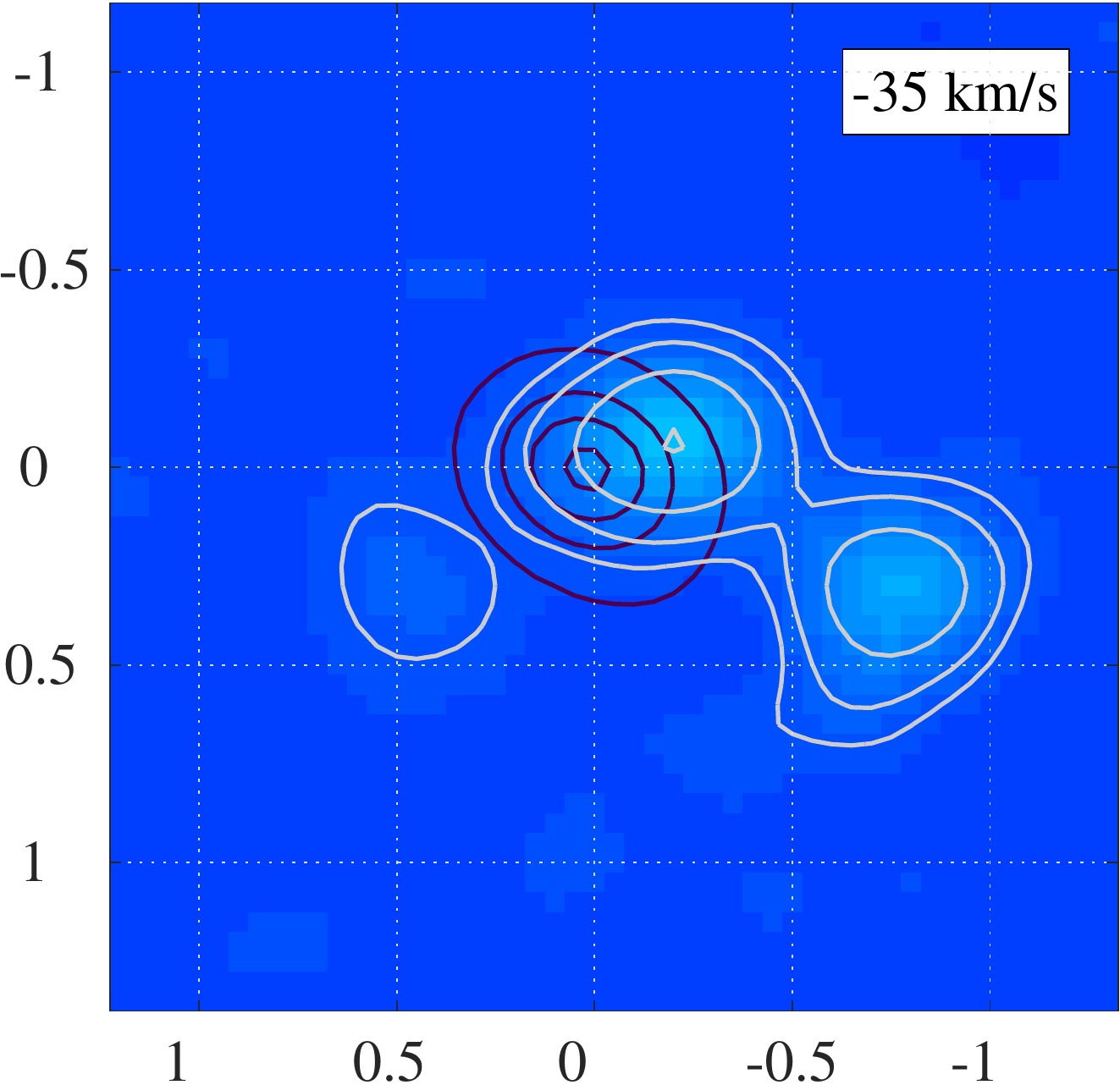}
   \includegraphics[width=4.43cm]{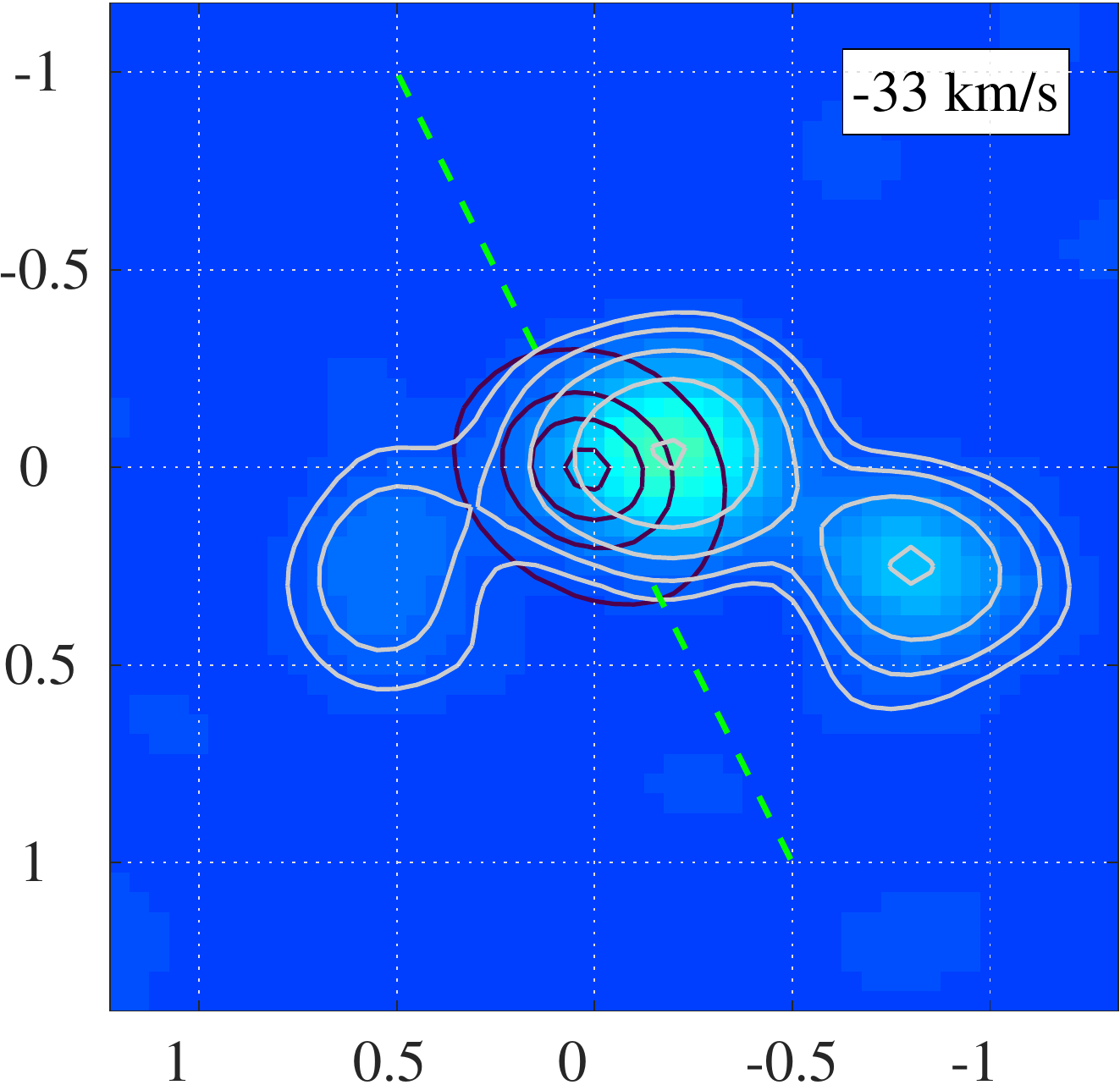}
   \includegraphics[width=4.43cm]{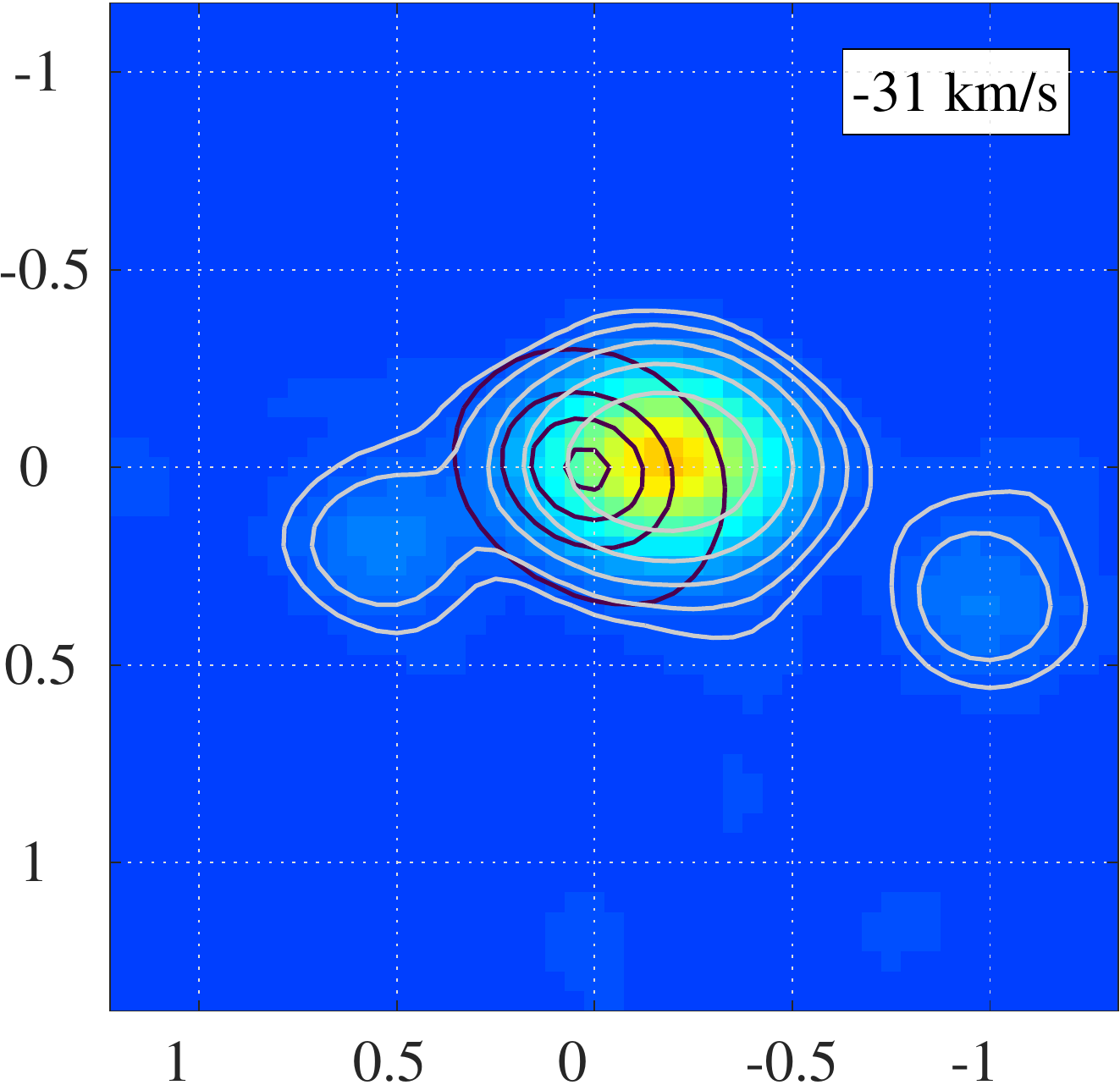}
   \includegraphics[width=4.75cm]{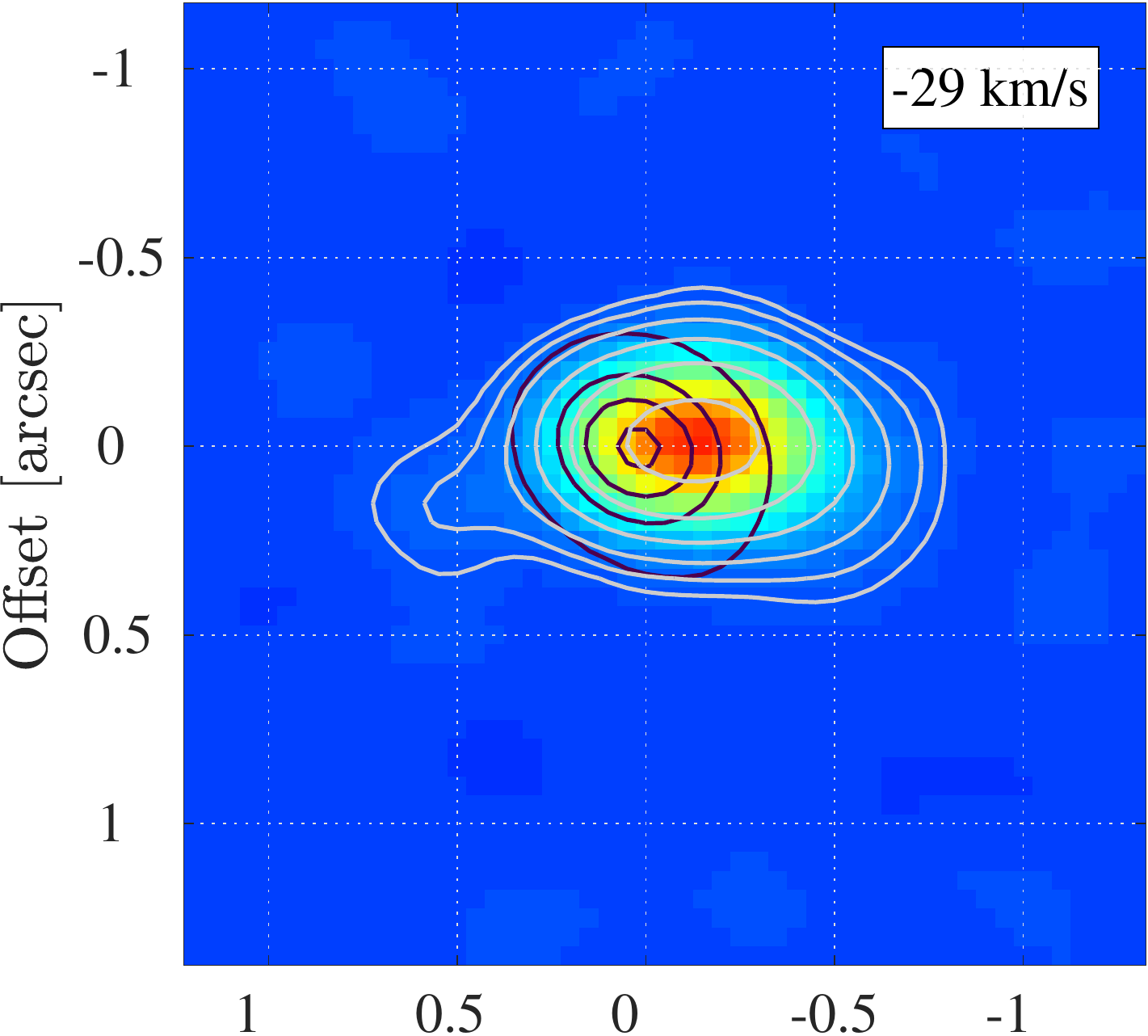}
   \includegraphics[width=4.43cm]{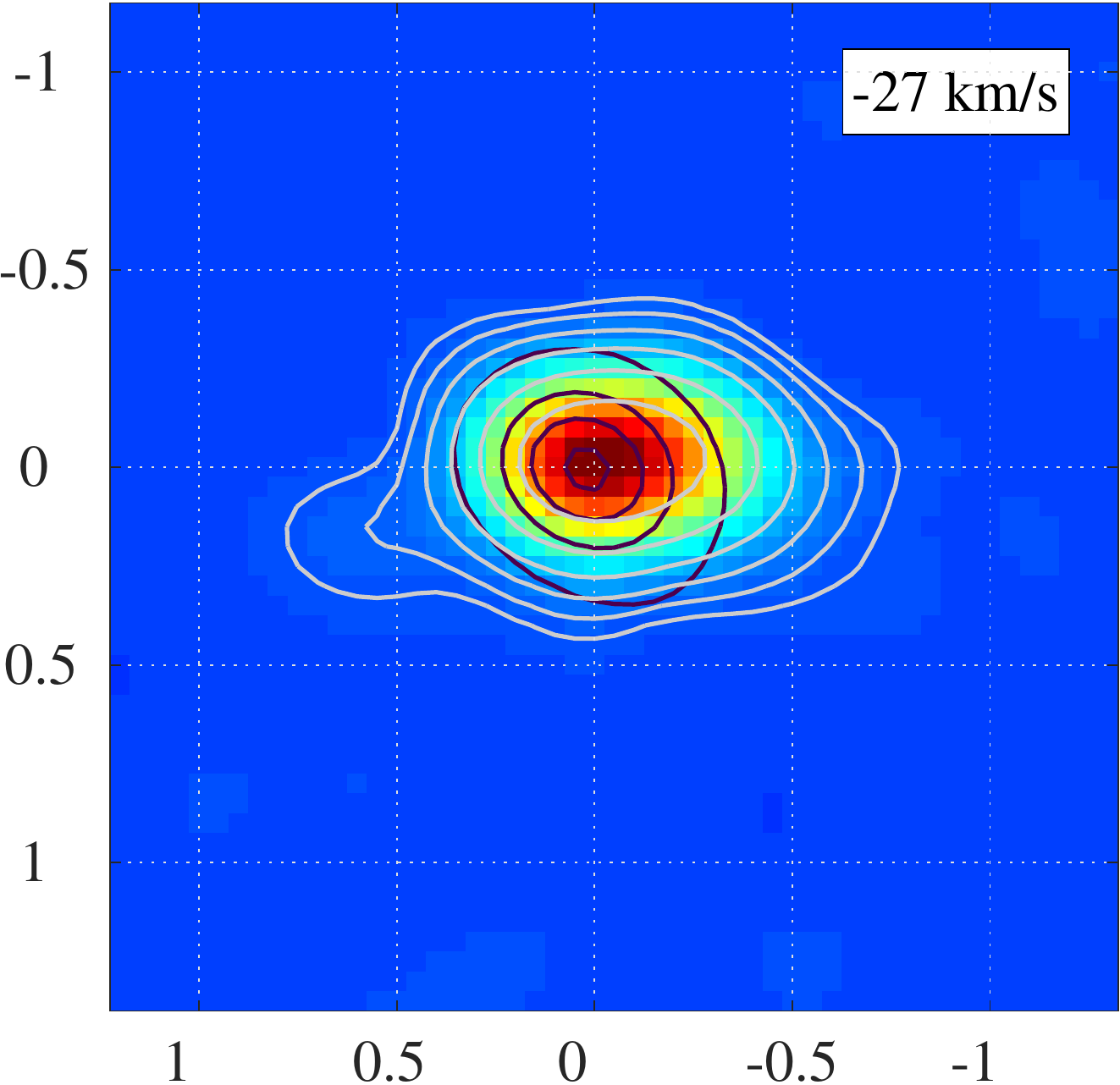}
   \includegraphics[width=4.43cm]{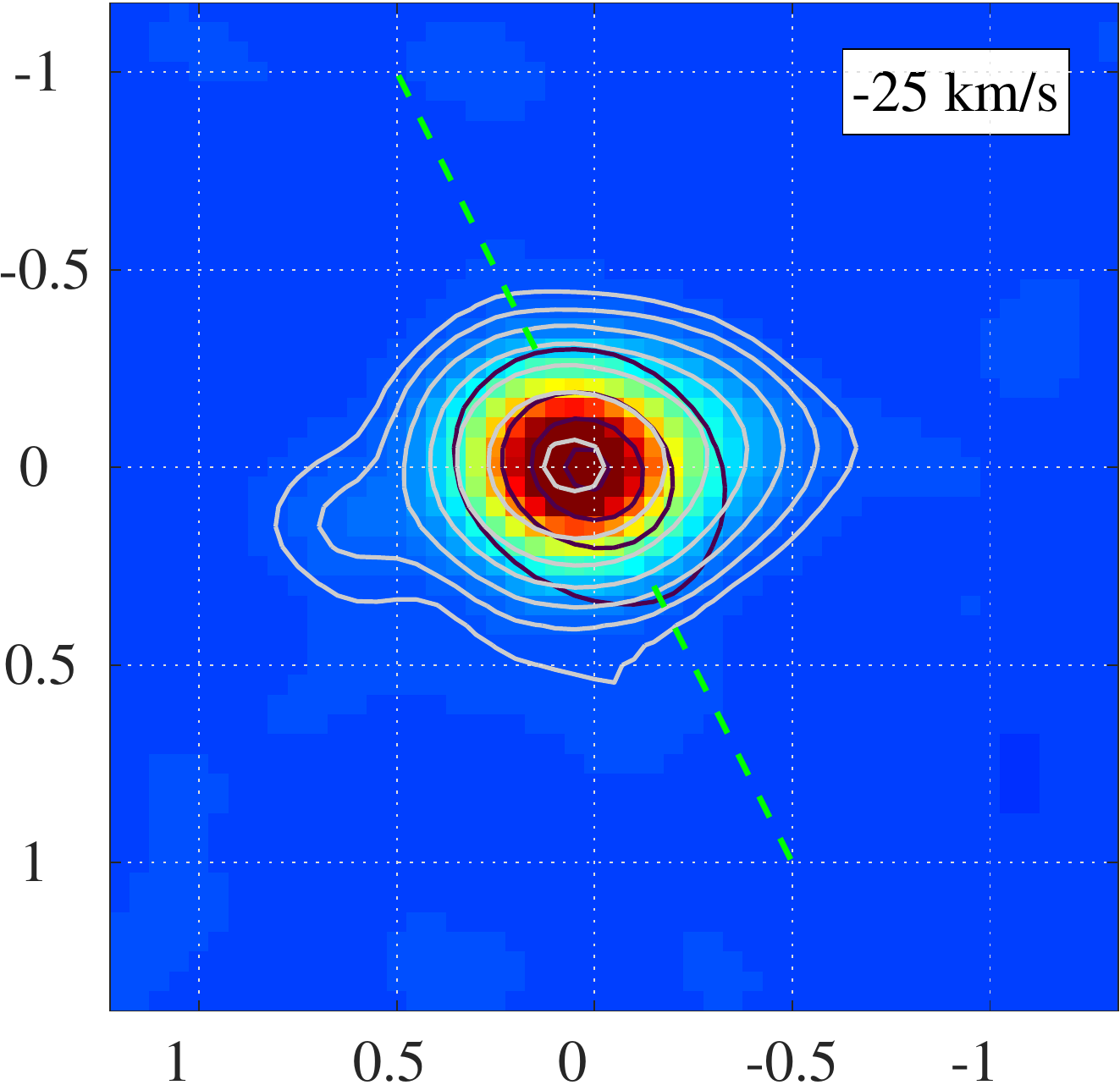}
   \includegraphics[width=4.43cm]{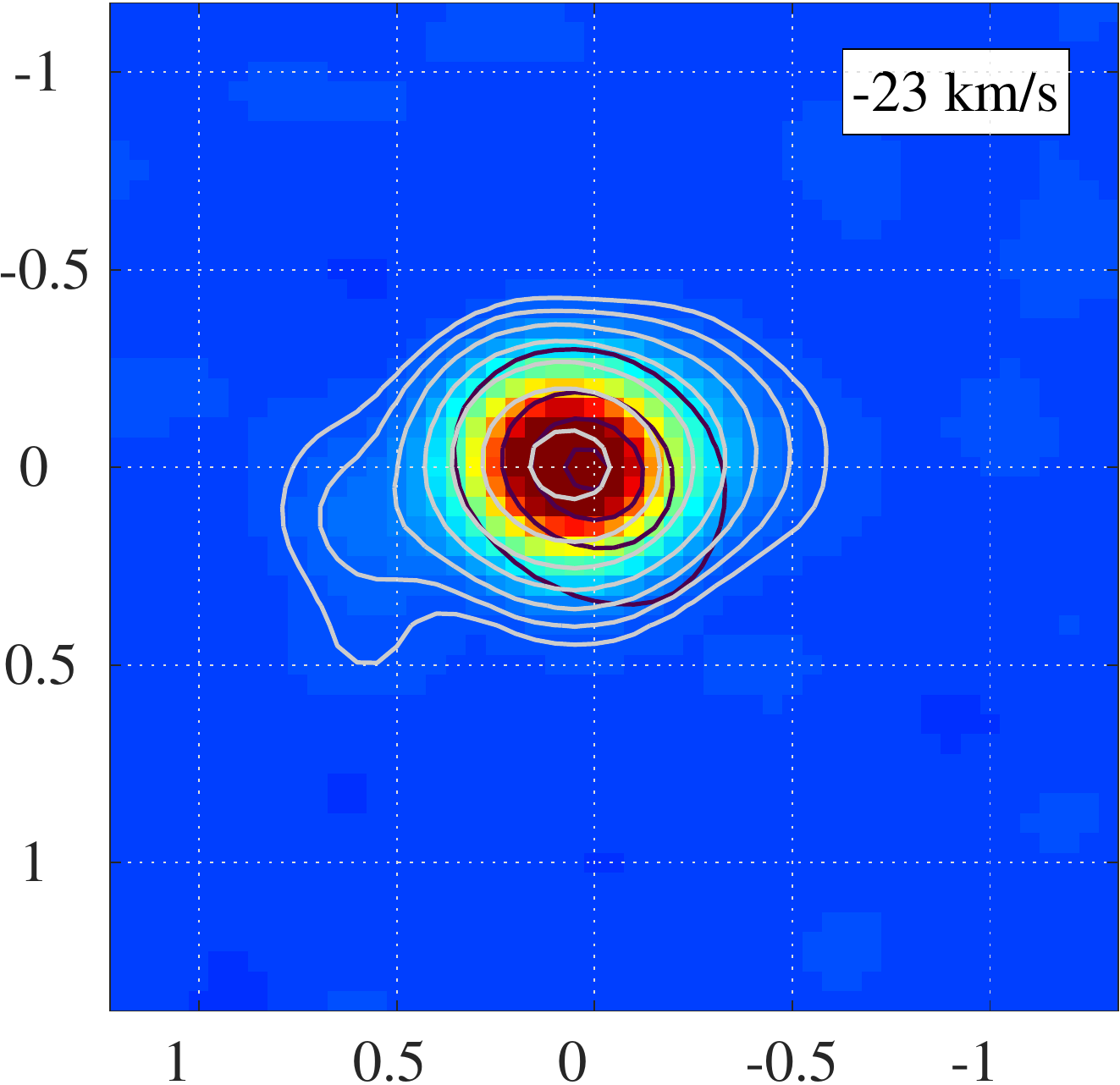}
   \includegraphics[width=4.75cm]{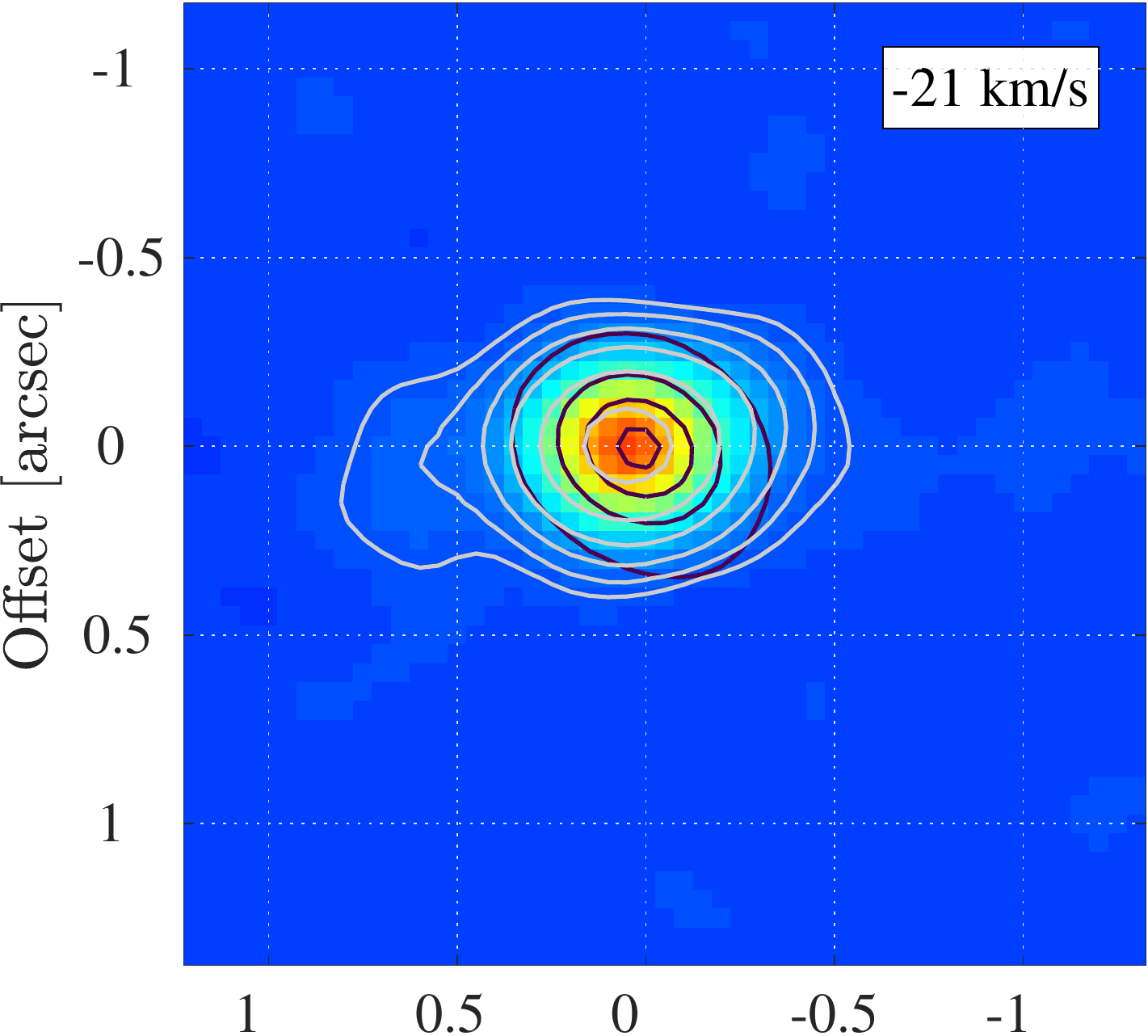}
   \includegraphics[width=4.43cm]{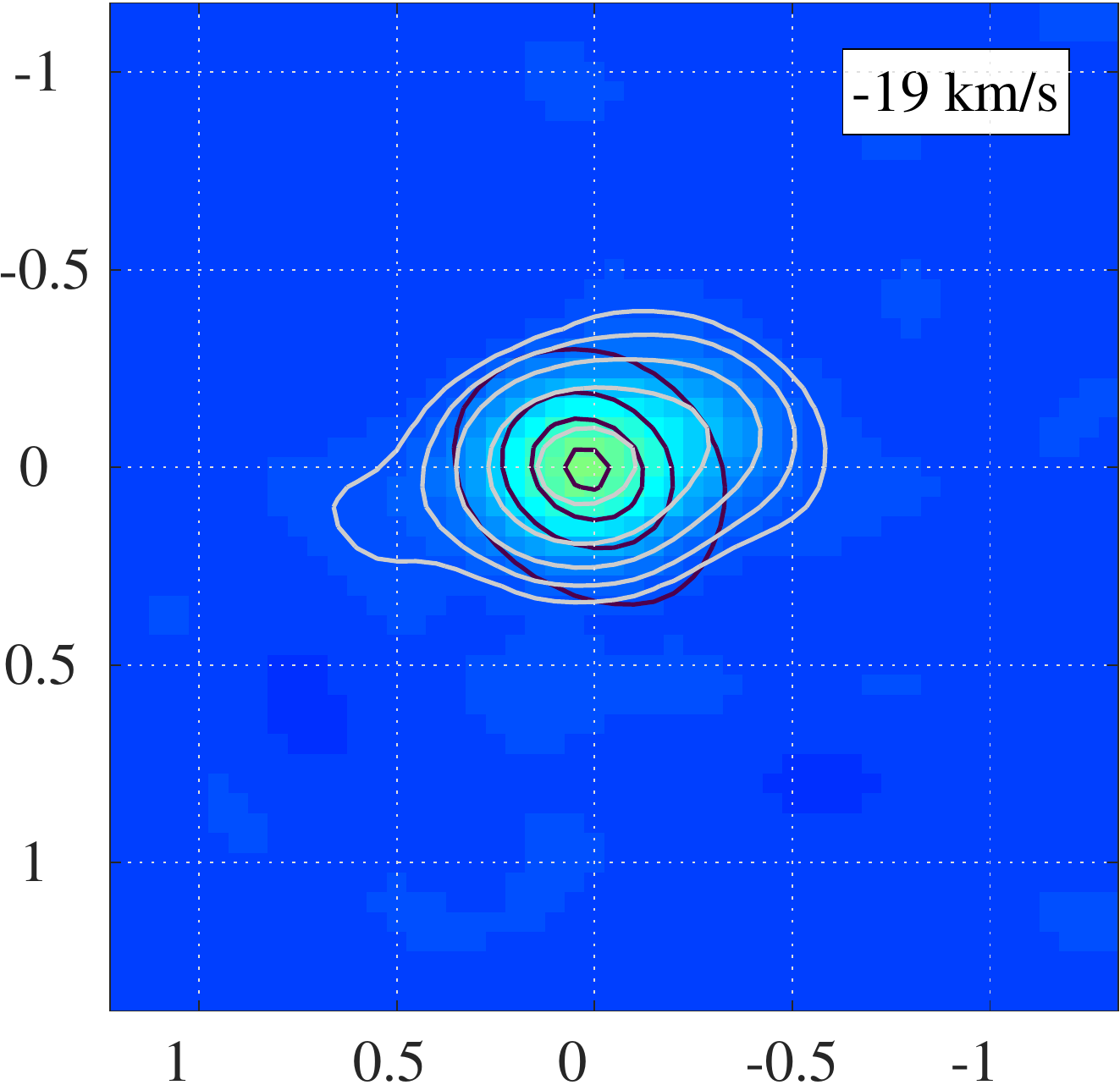}
   \includegraphics[width=4.43cm]{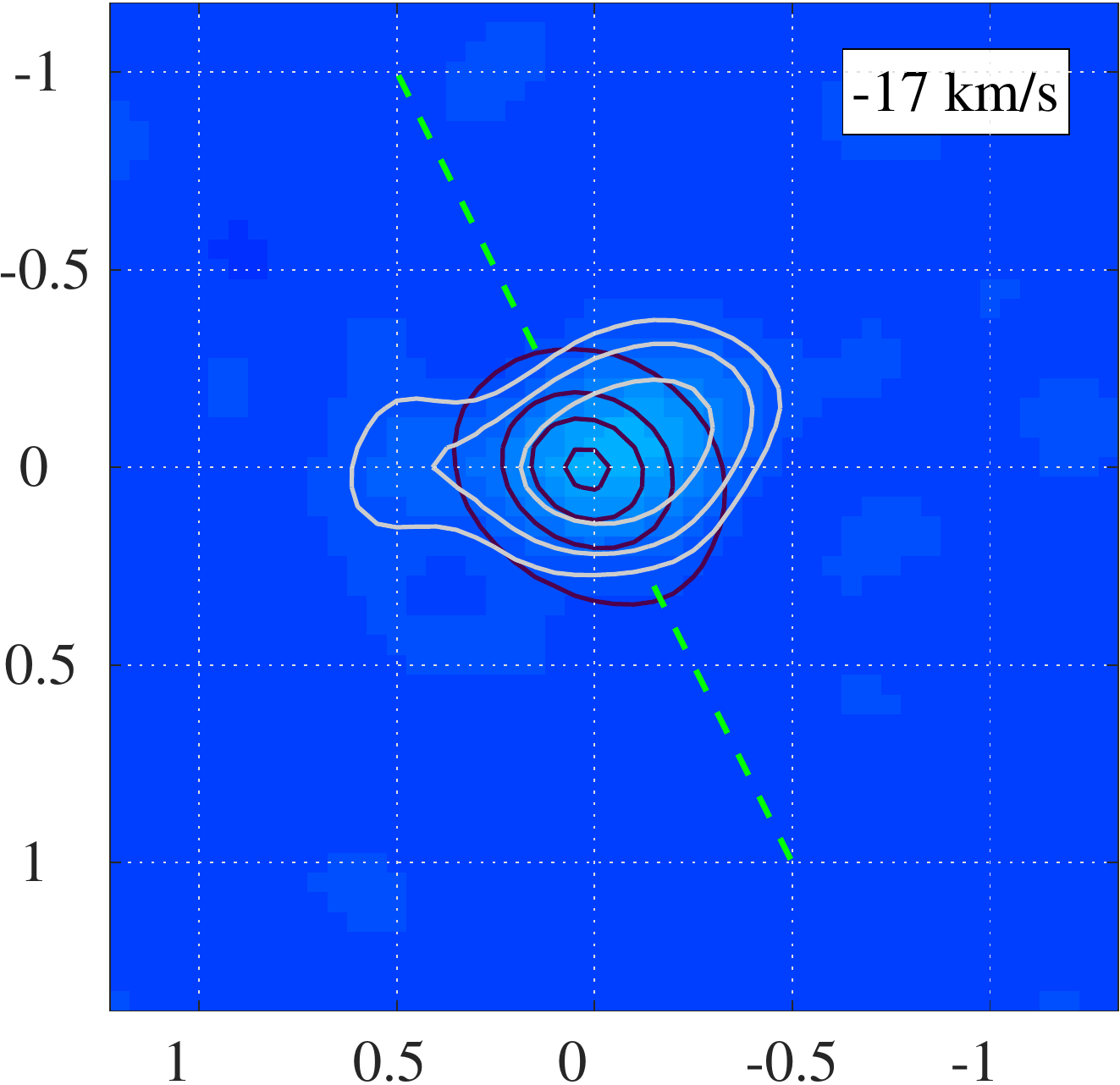}
   \includegraphics[width=4.43cm]{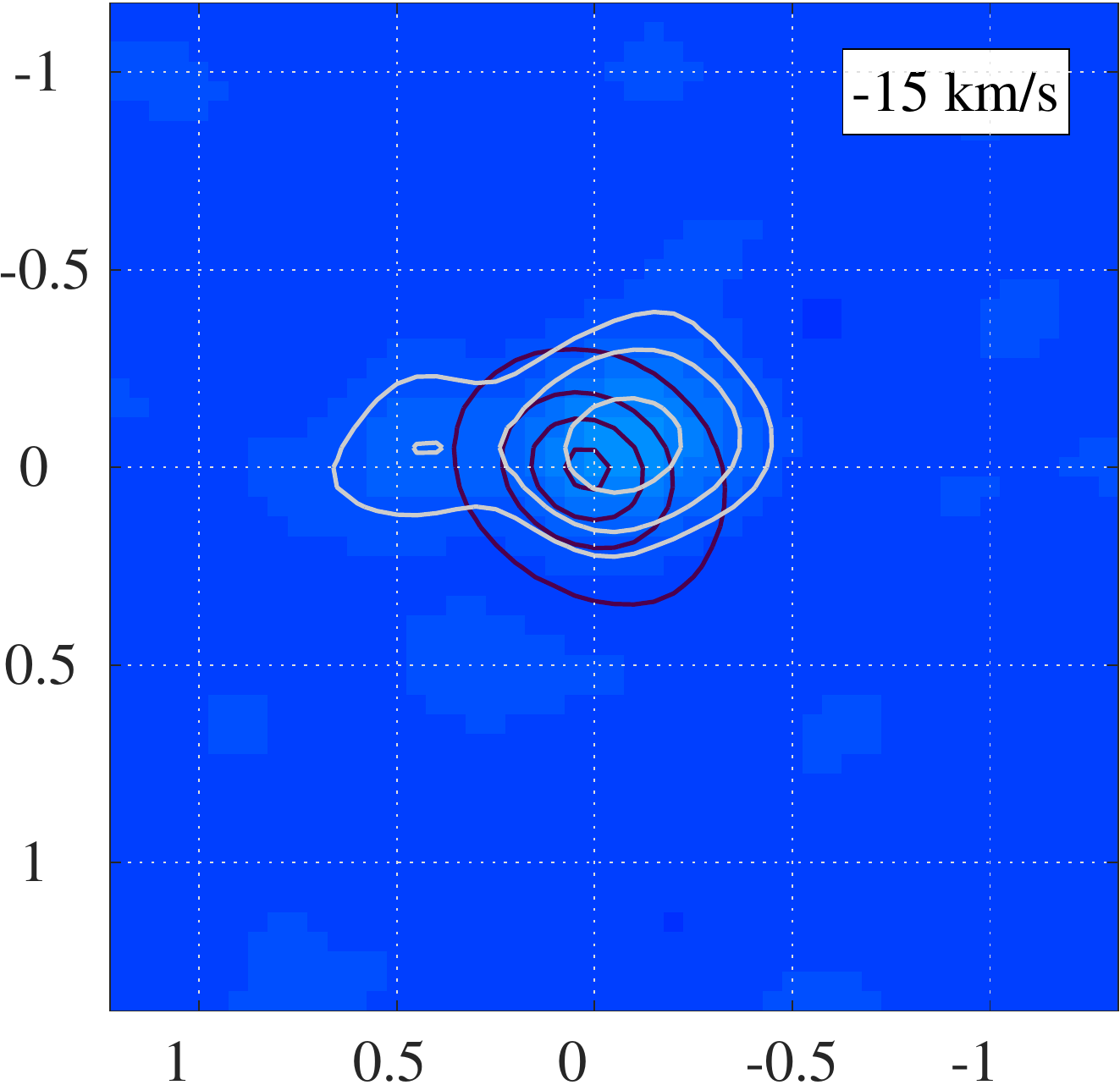}
   \includegraphics[width=4.75cm]{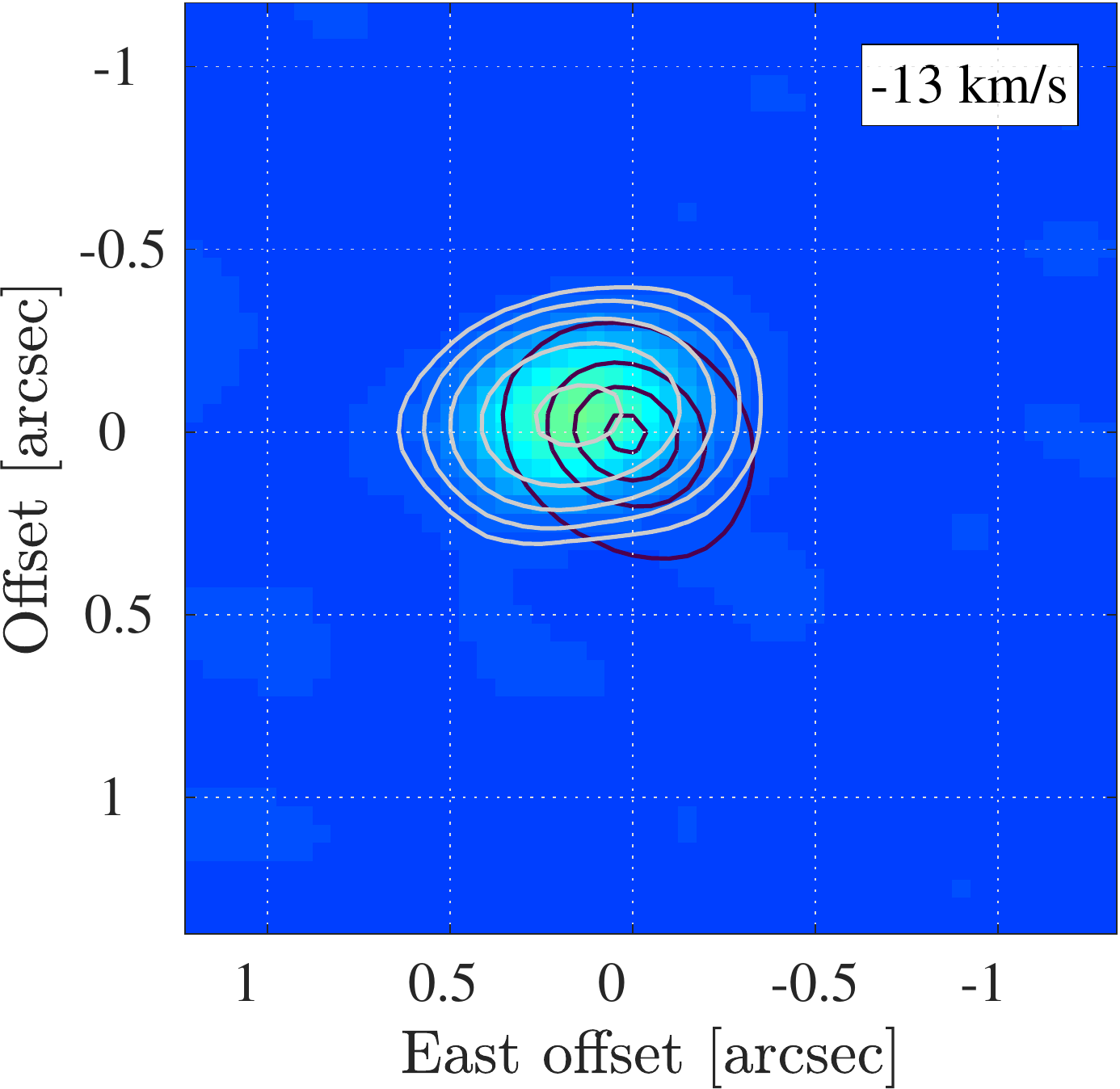}
   \includegraphics[width=4.43cm]{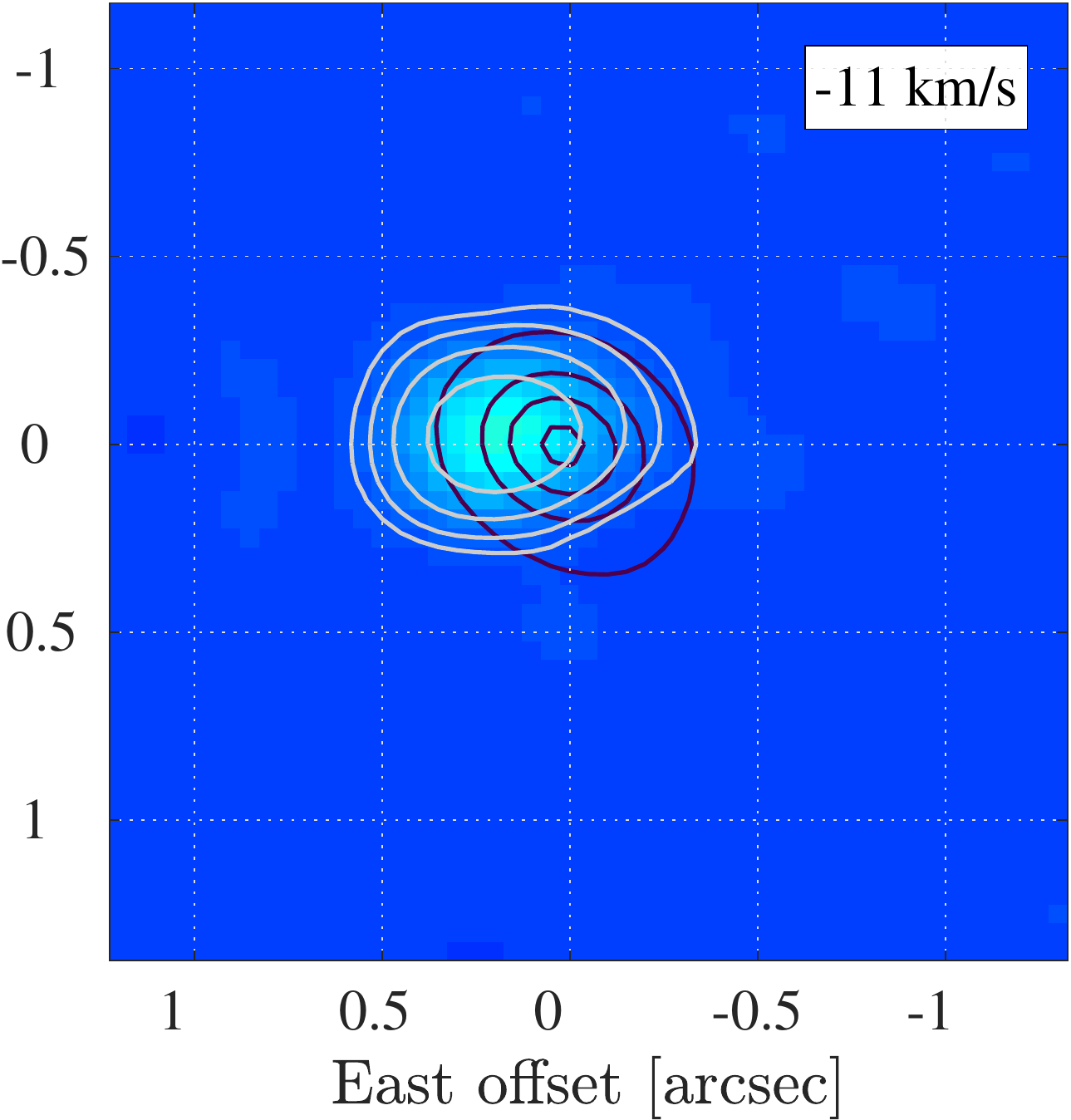}
   \includegraphics[width=4.43cm]{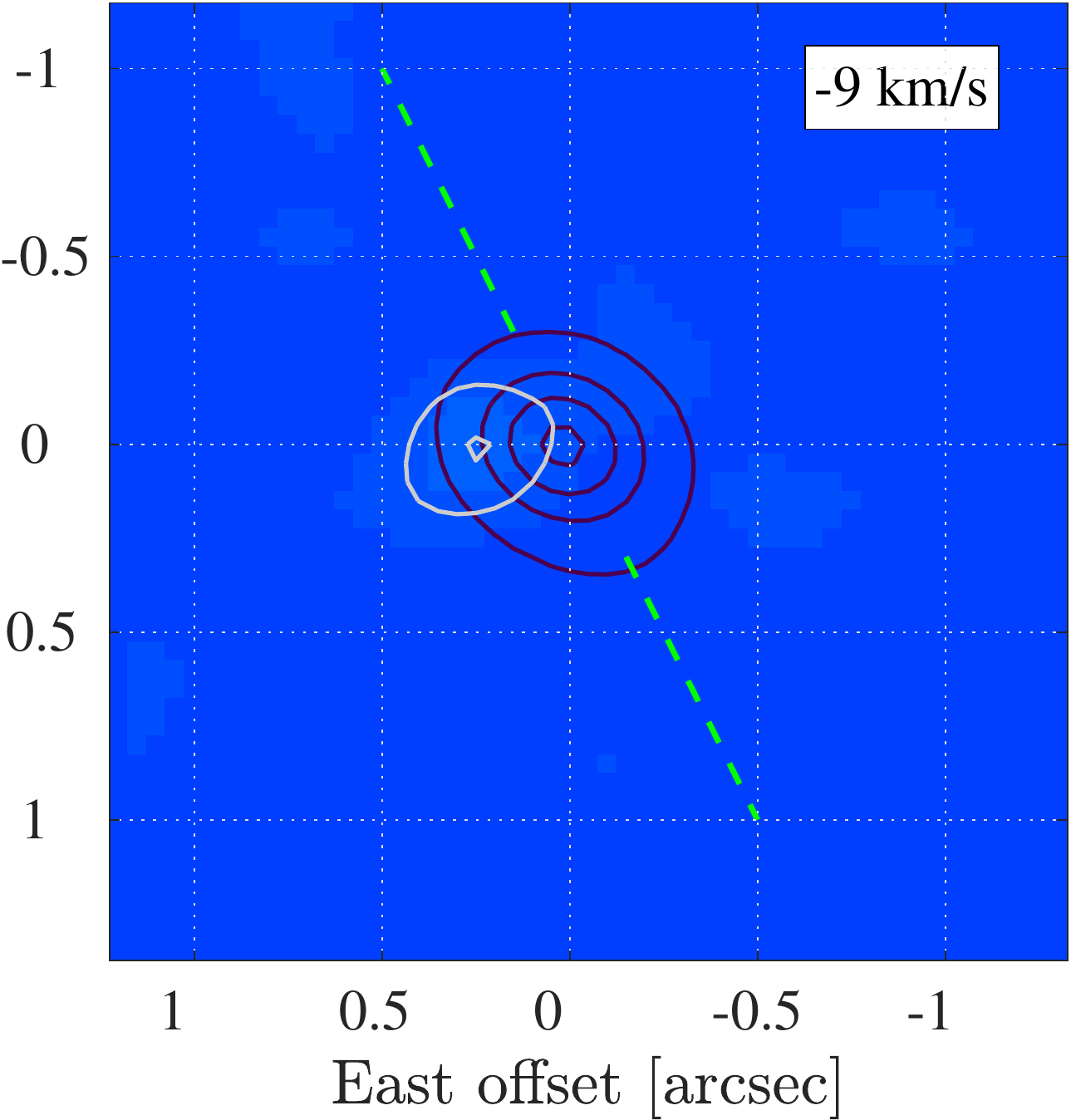}
   \includegraphics[width=4.43cm]{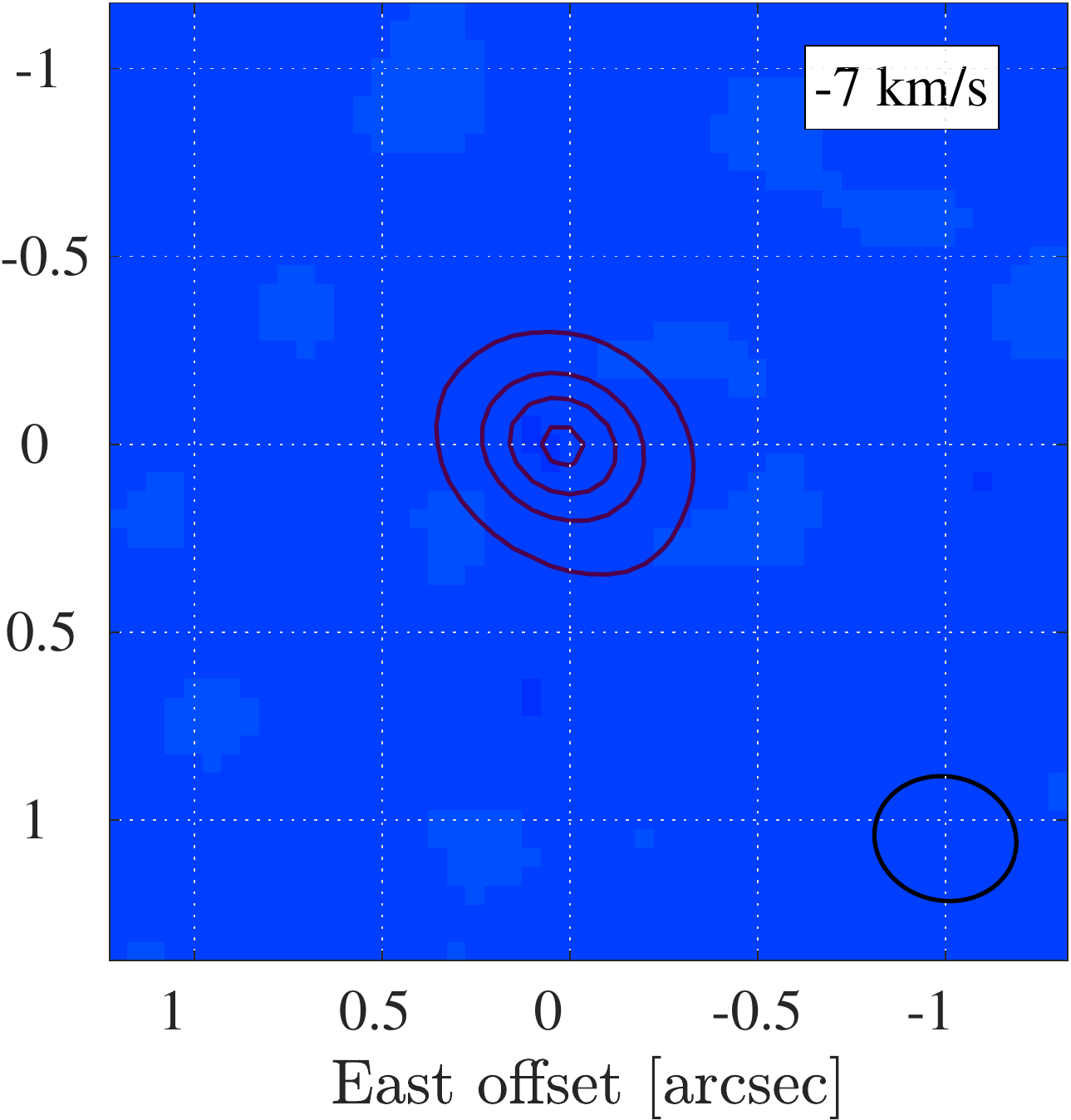}
   \caption{Images of the resolved $^{12}$CO(3-2) emission in different velocity channels (line-of-sight velocity indicated in the upper right corner). Thicker (maroon) contours centered at (0,0) show the unresolved continuum emission. Thinner (light gray) contours and the color scale show the line emission. The contours are drawn at 3, 6, 12, 24, etc. times the rms noise level in the emission-free channels. The images show the three-peaked, extended emission in the blueshifted channels, and the movement of the emission peak from west to east when moving from blue to red velocities. The (green) dashed line in the third-column images shows the direction of the jet \citep{schmetal17}.}
   \label{linemap}%
   \end{figure*}

%-------------------------------------------------------------------

\section{Data analysis}
\label{data}
The CO gas distribution will be affected by the dynamics of the system, but also by the hard radiation sources, that is, the white dwarf companion and emission from the jet, which can potentially photodissociate the CO molecules. In this section we describe the CO emission distribution and give possible interpretations for the underlying dynamical evolution. 

\begin{figure*}
\centering
\includegraphics[width=5.7cm]{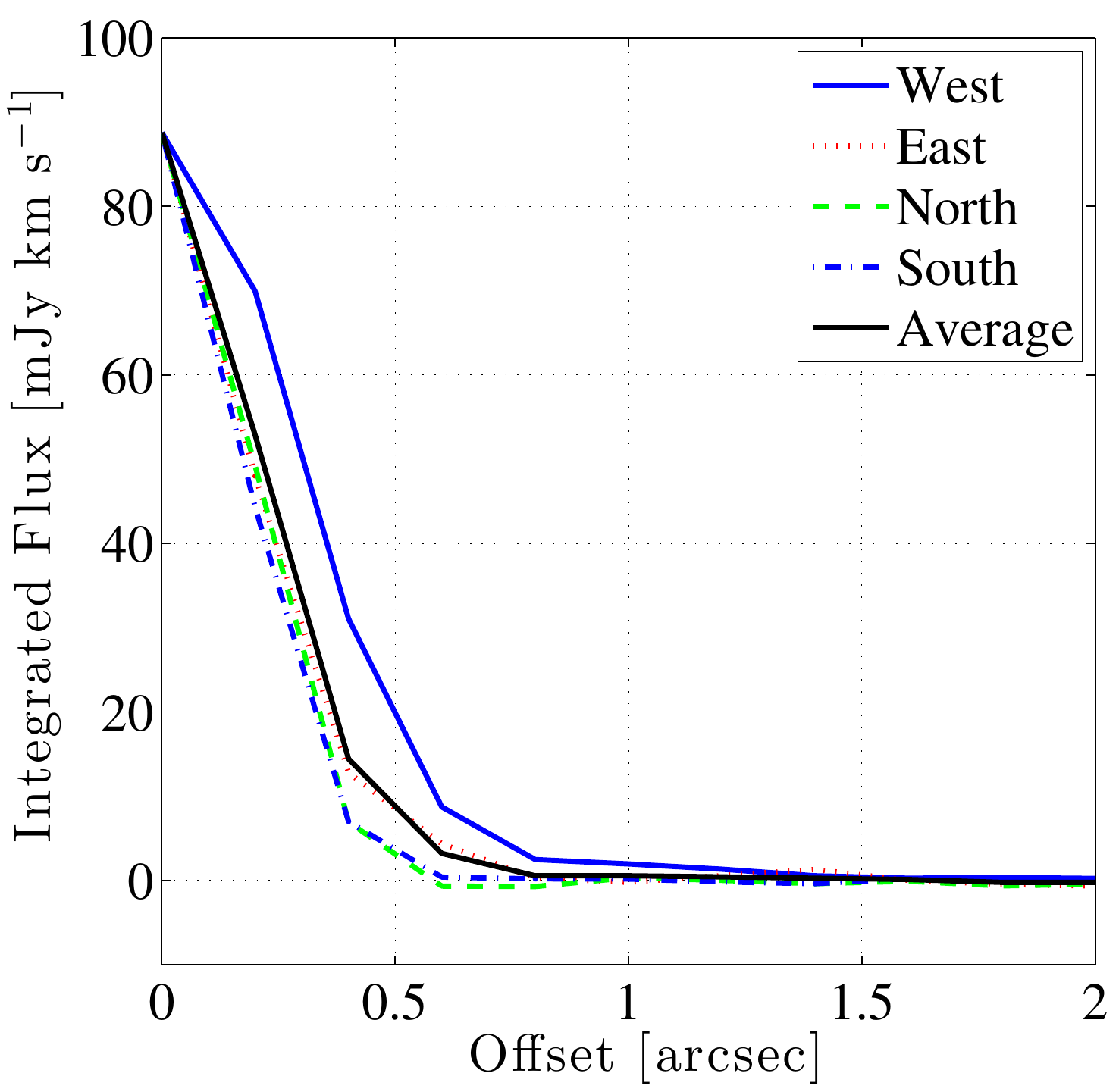}
\includegraphics[width=5.5cm]{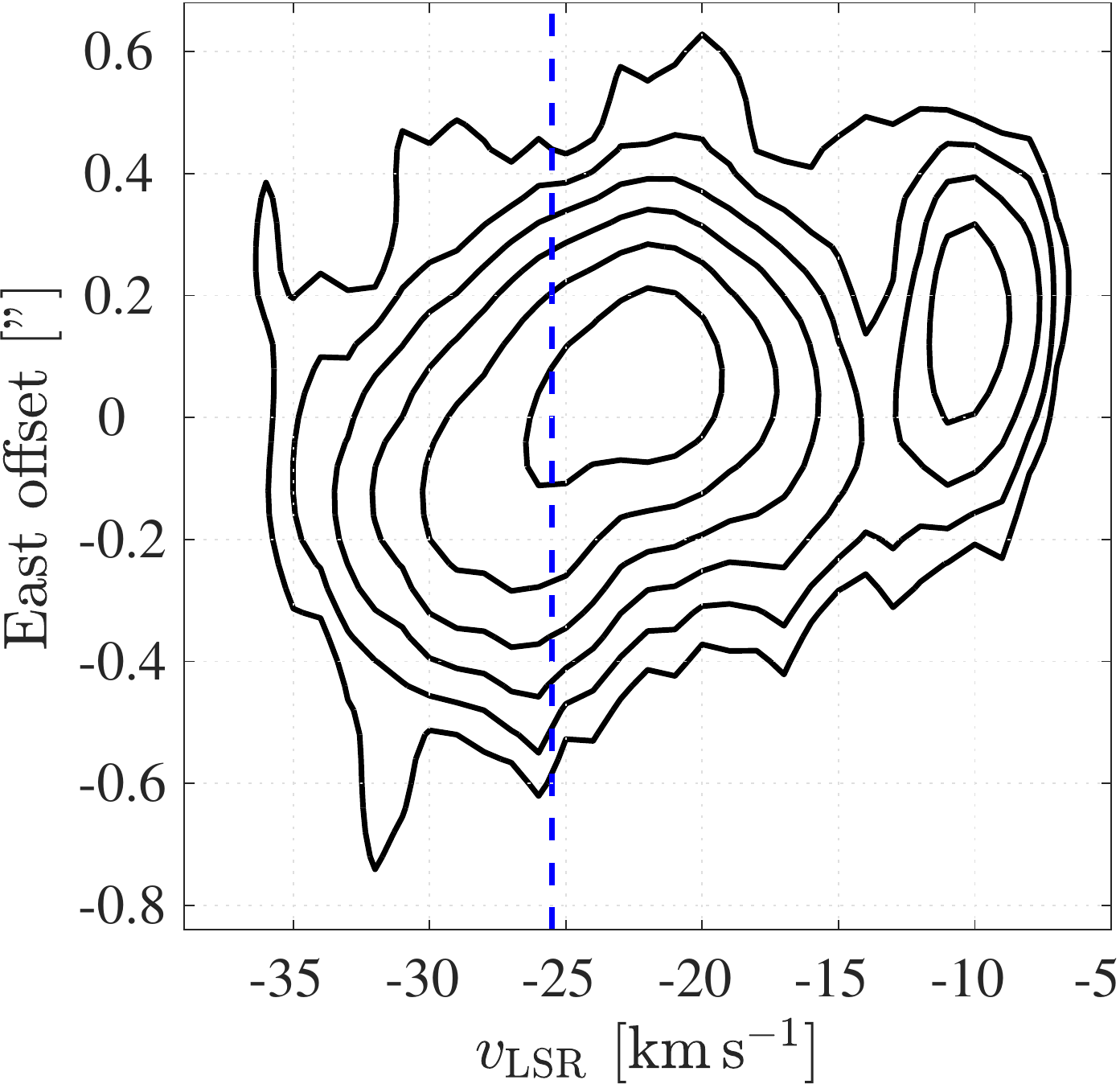}
\includegraphics[width=5.7cm]{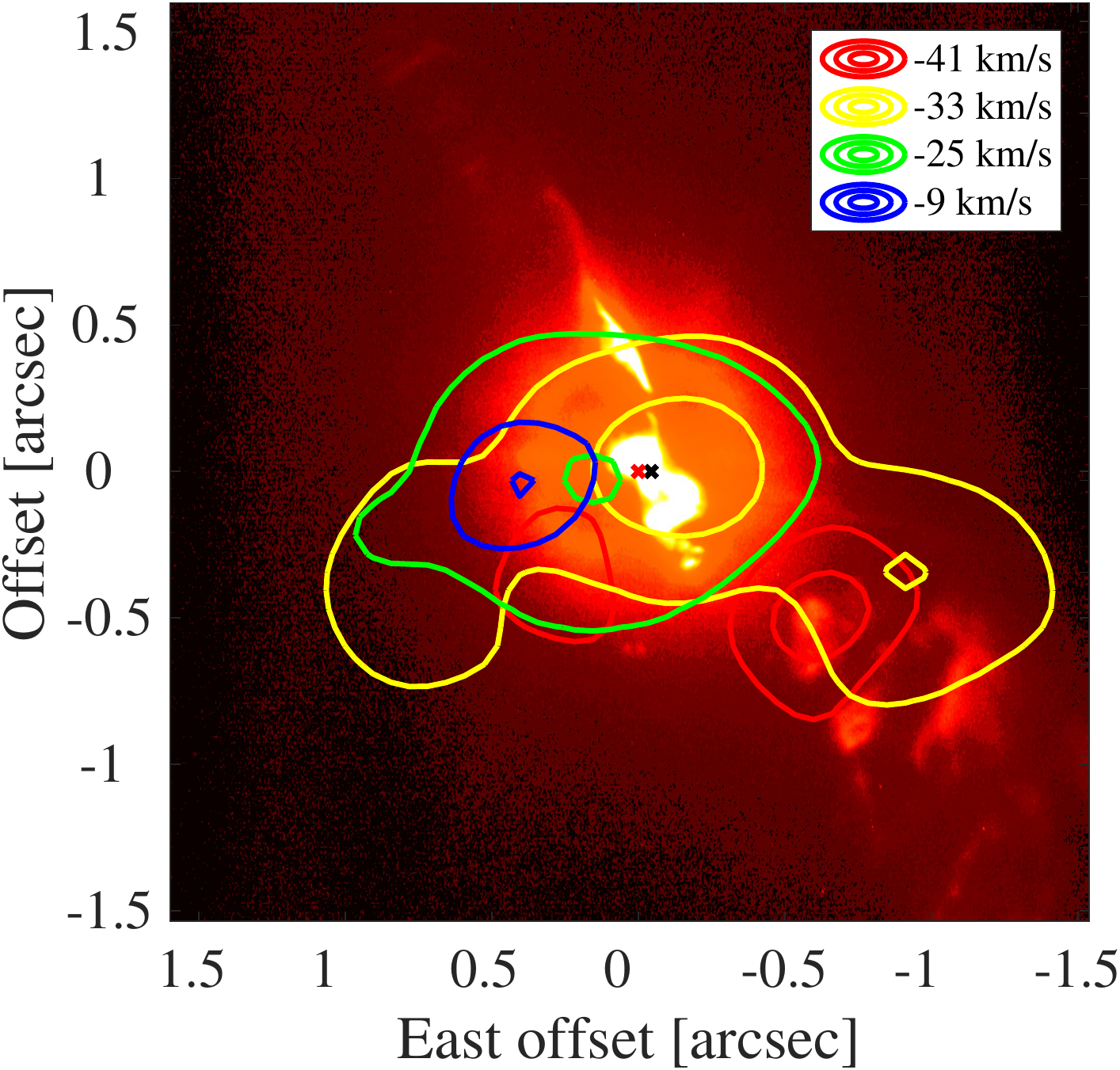}
\caption{{\bf Left:}  CO(3-2) line brightness measured every 0\farcs2  east, west, north, and south from the brightest pixel. {\bf Middle:} PV diagram of the CO(3-2) emission measured across a thin slit in the east-west directions. The contours are the same as in Fig.~\ref{linemap}. The dashed blue line marks the average radial velocity of the Mira star according to \citet{grommiko09}. {\bf Right:} Contours of the different main features at different velocities (shown in the legend) of the CO gas distribution overlaid on the H$\alpha$~image from \citet{schmetal17}, showing the jet and the resolved binary pair. The AGB star and white dwarf are marked by a red and a blue cross, respectively.}
\label{bright}
\end{figure*}

\subsection{Average mass-loss rate}
\label{ave-mdot}
The ALMA data presented in this paper provide the first resolved maps of the CO emitting region around R~Aqr, and it is useful to estimate an outer radius in order to update previous estimates of the average mass-loss rate from the AGB star. The CO emitting shell is not spherically symmetric (Fig.~\ref{linemap}), but an average outer radius can be constrained by measuring the brightness in the images along different directions. The size of the emitting region in Fig.~\ref{linemap} is substantially smaller than the maximum recoverable scale of the observations (see Sect.~\ref{obs}) and is therefore limited by the excitation conditions for the CO(3-2) line.

The CO(3-2) line brightness distribution (Fig.~\ref{bright}) was measured by first smoothing the whole image cube to 0\farcs4 resolution using the CASA task {\tt imsmooth}. The line emission was then measured every 0\farcs2 (and integrated) north, south, east, and west from the brightest pixel out to 2.0\arcsec (Fig.~\ref{bright}). The east-west elongation seen in Fig.~\ref{linemap} is again obvious. The brightness profile was averaged over the four directions and the outer radius (where the CO(3-2) line emission is no longer detected) is 0\farcs8, which corresponds to about 170\,AU at 218\,pc. 

\citet{bujaetal10} estimated the mass-loss rate of R~Aqr by fitting the CO(2-1) line observed at the IRAM 30\,m telescope. By assuming that the size of the emitting region is 2$\times$10$^{14}$\,cm and a CO fractional abundance of 5$\times$10$^{-4}$, they estimated a mass-loss rate of $\dot{M}$=9$\times$10$^{-6}$\,M$_{\odot}$\,yr$^{-1}$. From previous radiative transfer modeling results \citep[e.g.,][]{ramsetal08}, we estimate that the outer radius (defined as the e-folding radius, $R_{\rm{e}}$) will scale as $\sim$1.4 times the CO(3-2) half-brightness radius, and that the mass-loss rate will scale as $R_{\rm{e}}^{-1.82}$. With the size derived from the ALMA observations, the mass-loss rate will then be on the order of $\dot{M}$=2$\times$10$^{-7}$\,M$_{\odot}$\,yr$^{-1}$. It should again be noted that this gives an average value during the creation of the CO-emitting CSE and is not necessarily representative of the present-day mass-loss rate from the Mira star in the R~Aqr system. The standard model used to estimate mass-loss rates from CO lines \citep[e.g.,][]{schoolof01,debeetal10} is build for a spherically symmetric, isotropically expanding, homogeneous CSE, created by the stellar wind during thousands of years. Previous models based on this standard setup have guided the assumptions made for the mass-loss-rate estimate that is given here. However, what is generally given as an observed estimate of the mass-loss rate, that is, the average mass-loss rate during the creation of a several thousand AU, spherically symmetric CSE, is not really applicable to a case like the rather young ($\sim$100\,AU) CO envelope around the R Aqr system, which consists of several different dynamical and morphologically complex components, probably with varying CO/H2.

\subsection{CO gas distribution and line emission}
\label{dist}
Figure~\ref{linemap} shows the channel maps at 2\,km\,s$^{-1}$ spectral resolution. The CO gas envelope is confined and an even finer spatial resolution would certainly help in interpreting the circumstellar emission distribution. However, some structure is resolved even with this data set: At blueshifted velocities, emission at 3$\sigma$ first appears at $-43$\,km\,s$^{-1}$ at 0\farcs3 south of the peak of the continuum emission. From $-41$ to $-39$\,km\,s$^{-1}$, the emission is still to the south and appears at two positions on either side of the continuum peak. At $-37$\,km\,s$^{-1}$, a third emission component slightly to the northwest of the continuum peak appears. From $-35$ to $-31$\,km\,s$^{-1}$ , the three components remain and the center component increases in brightness. After $-31$\,km\,s$^{-1}$ , only the center component is apparent and moves to the east, while the brightness eventually decreases and reaches a minimum at $-15$\,km\,s$^{-1}$. From $-13$ to $-9$\,km\,s$^{-1}$ , the emission brightens again and appears to the east of the continuum peak. At velocities above $-7$\,km\,s$^{-1}$ , no emission is detected. The brightness distribution is also reflected in the $^{12}$CO(3-2) line profile shown in Fig.~\ref{COline}. The $^{12}$CO line emission starts at $-43$\,km\,s$^{-1}$, peaks at $-24$\,km\,s$^{-1}$, has a minimum at $-15$\,km\,s$^{-1}$, another dimmer peak at $-12$\,km\,s$^{-1}$, and disappears beyond $-7$\,km\,s$^{-1}$. 

\begin{figure*}
\centering
\includegraphics[width=18cm]{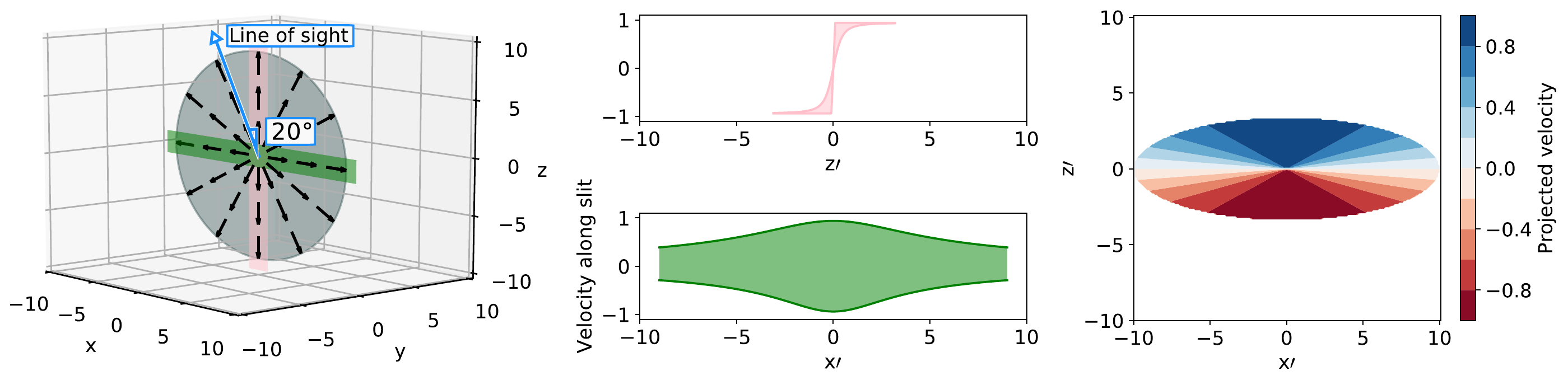}
\includegraphics[width=18cm]{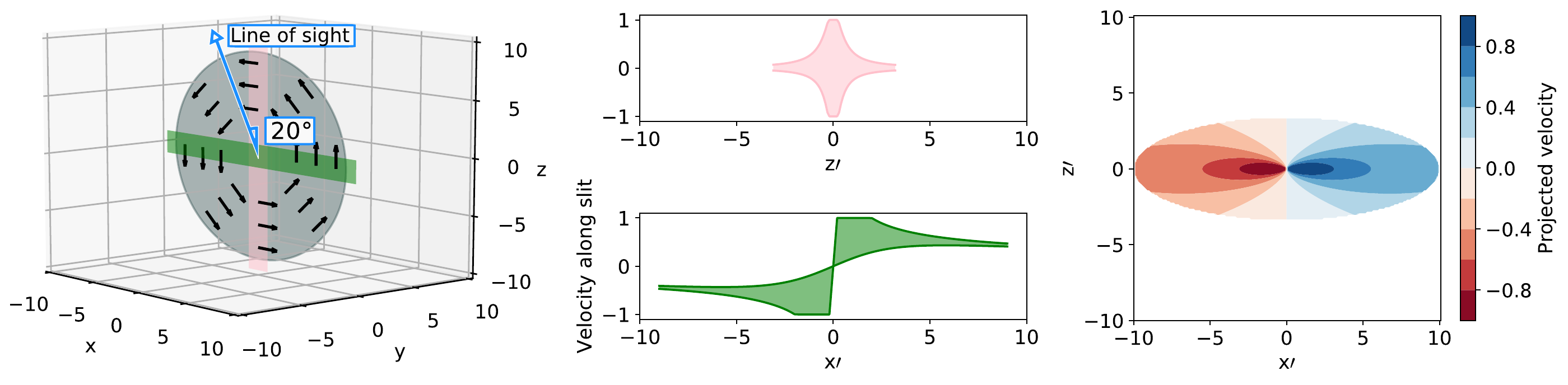}
\caption{Demonstrating the degeneracy between observations of rotation and expansion in a thin layer. {\bf Left:} System orientation. {\bf Middle:} Velocity-position diagrams generated across the pink and green region marked in the left figure. {\bf Right:} Velocity distribution across the sky when the system is viewed at an angle 20$^{\circ}$ below edge on, i.e., the angle between the line of sight and the orbital plane in the R~Aqr system.}
\label{pv}
\end{figure*}

\subsection{Wind variations along the orbit}
\label{wind}
Figure~\ref{COline} (right) shows the relative positions of the binary pair along the orbit as derived by \citet{grommiko09}. We have marked the positions at the latest periastron passage (green stars) and the positions at the time of the ALMA observation. The stars are currently approaching the next periastron passage, which will take place in January 2023. According to the results of \citet{grommiko09}, the Mira star reaches its maximum receding radial velocity of approximately $-21$\,km\,s$^{-1}$ at 0.8 orbital phase, that is, just before the ALMA observation. It reaches its maximum approaching velocity of approximately $-29$\,km\,s$^{-1}$ at 0.1 orbital phase. This is marked by thick black arrows in Fig.~\ref{COline} (right).

A possible explanation for the three-part blueshifted emission components peaking around $v_{\rm{LSR}}=-33$\,km\,s$^{-1}$ (also shown as yellow contours in Fig.~\ref{bright}, right) is that the mass-loss rate has varied as the Mira star moves along its orbit. A relatively short-term (on the order of tens of years) increase in the mass-loss rate along the equator plane would create an expanding ring around the Mira star. In this scenario, the two southern components from $-41$ to $-31$\,km\,s$^{-1}$ would correspond to the central cross-section of the ring. If the mass-loss-rate increase occurred when the Mira star was at maximum approach, the ring should be centered around the channel at $-29$\,km\,s$^{-1}$. If the emission at $-43$\,km\,s$^{-1}$ is the front side of the ring, it is currently expanding at $14$\,km\,s$^{-1}$. The minimum and maximum width of the ring in the channel at $-33$\,km\,s$^{-1}$ is 0.9\arcsec and 2\arcsec, respectively. Assuming a constant expansion velocity, this means that the ring would have been created from approximately 70 to 30 years ago, at which time the Mira star passes from J1 to S (see Fig.~\ref{COline}, right) anticlockwise, covering almost the full orbit. 

The north-south offset of the two southern features does not exactly fit the orbit according to \citet{grommiko09}. They derived an orbit inclination of 70$^{\circ}$ from the pole and a semimajor axis of the Mira orbit on the order of 6\,AU. At maximum separation, this corresponds to a north-south offset of 0.06\,AU, assuming a distance of 218\,pc. However, given the uncertainties in the orbit calculation, the distance, and the data, it is still plausible that the north-south offset is due to the ring being created at a different position along the orbit.

In the same scenario, the center component, visible from $-37$ to $-9$\,km\,s$^{-1}$ (also shown as green contours in Fig.~\ref{bright}, right), is the CSE created by the more recent mass loss from the Mira star. With the size estimate from Sect.~\ref{ave-mdot} of an e-folding radius of about 110\,AU and an expansion velocity of 14\,km\,s$^{-1}$, the dynamical timescale for the creation of the CSE is about 40\,yr, which is about one orbital period \citep{grommiko09}. According to the right panel in Fig.~\ref{COline}, movement from west to east along the orbit (anticlockwise) should correspond to a blueshift of the velocity. The shift from west to east with increasing velocity seen in the channel maps (and by comparing the position of the different velocity contours in Fig.~\ref{bright}, right) might be explained only by the movement along the ($\sim$10\,AU=0\farcs05 wide) orbit.  This shift, as well as the brightness dip around $-16$\,km\,s$^{-1}$, could also be explained by the hot binary companion shaping arcs (with material flowing along them), as has been seen in several recent observations of binary AGB stars \citep[e.g.,][]{ramsetal14,cernetal15,kimetal17}. Upcoming better spatial resolution observations will be able to confirm whether this is the case in R~Aqr as well.

\subsection{Can rotation be detected?}
A rotating, circumbinary disk, such as that which has been found around the post-AGB objects the Red Rectangle and AC~Her, for instance, would not be unexpected in a system like R~Aqr (see also Sect.~\ref{context}). Figure~\ref{pv} shows the expected emission distribution from an infinitesimally thin, uniform, optically thin round plate viewed from an angle of 20$^{\circ}$ below edge-on (the expected viewing angle of a disk aligned with the orbital plane in the R~Aqr system). The upper panel shows the plate with a radial velocity field, and the lower panel shows a disk with Keplerian rotation. The middle figures show the velocity-position diagram across a north-south (pink) and an east-west (green) slit. The right figures show the projected velocities of the plate across the sky, that is, the first moment of the channel maps. The middle figure of Fig.~\ref{pv} shows that the idealized case of a rotating disk viewed edge-on will result in a characteristic bow-tie shape in a position-velocity (PV) diagram when produced along the disk. It is also shown that rotational movement and radial expansion can give similar signatures in PV diagrams depending on the orientation of the system (see uppermost and lowermost panel) and additional information about the orientation of the system becomes necessary to interpret the PV diagram. The rotating disks found around post-AGB stars \citep{bujaetal13} are an order of magnitude larger than the extent of the molecular emission around R~Aqr.  In the Red Rectangle \citep[e.g.,][and references therein]{bujaetal16}, for example, rotation and expansion are simultaneously present, but to be able map the exact dynamics, as for some post-AGB stars, requires very well resolved observations. This is more difficult to acquire for the less extended emission or CSEs of AGB stars. 

The middle panel in Figure~\ref{bright} shows a PV diagram of the CO(3-2) emission from R~Aqr produced using a thin slit along the east-west direction. We note again that the spatial resolution is about 0\farcs3. The dashed, vertical line marks the average $v_{\rm{LSR}}$ of the Mira star according to \citet{grommiko09}. The emission distribution from $-35$ to $-15$\,km\,s$^{-1}$ is consistent with a rotating structure where the eastern side moves away from us and the western side moves toward us (see the lowest middle panel of Fig.~\ref{pv}), that is, along the orbital motion of the stars \citep{grommiko09}. However, with the resolution of the data, we cannot with certainty determine whether rotation is present in the molecular gas around R~Aqr. The upcoming higher spatial resolution data mentioned above will be very valuable to distinguish the dynamics of the R~Aqr system.

\subsection{Line emission from molecules other than $^{12}$CO}
The other lines (Table~\ref{lines} and Fig.~\ref{COline}, middle) are approximately as wide as the $^{12}$CO(3-2) line, but show absorption between -17 and 0\,km\,s$^{-1}$. While the second $^{12}$CO(3-2) emission peak is at $-12$\,km\,s$^{-1}$, the absorption dip in the other lines is shifted from -13 to -10\,km\,s$^{-1}$ with the strength of the line. The shapes of the $^{12}$CO  line and of the other lines are indicative of complex dynamics, different from the "normal" AGB wind. \citet{khouetal16} used a simple two-component model (warm molecular layer+outflow) to reproduce the inverse P-Cygni profile seen also in the vibrationally excited ($v$=1) CO(3-2) line detected in the same data set. The model ignores any influence of or flow to the companion, and the inverse P-Cygni profile can be reproduced when including infalling material close to the star, that is, in the warm molecular layer. The same profile shape is seen in all lines, except for the $^{12}$CO(3-2) line (which comes from a more extended region), and is likely caused by similar gas kinematics close to the Mira star. The $^{12}$CO(3-2) line shape seems very similar to that of the $^{12}$CO(2-1) line from the IRAM 30m telescope presented in \citet{bujaetal10}.

%------------------------------------------------------------------

\section{Molecular emission in context}
\label{context}

The largest structure associated with the R~Aqr system is the bipolar nebula, where the $\sim$40\arcsec-semimajor-axis waist is clearly seen in H$\alpha$ images from the Very Large Telescope (VLT) \citep{liimetal18} and Hubble Space Telescope (HST) \citep{schmetal17}. The waist is extended in the east-west, perpendicular to the north-south bipolar nebula, and centered on the binary pair.  It appears to be expanding at the same inclination angle as the orbit, that is, at 70$^{\circ}$ from the north. The same structure is detected at centimeter wavelengths \citep{hollet87} and at 70 and (partly at) 160 $\mu$m with the Herschel/PACS instrument \citep{mayeetal13}.

A large cuved jet lies at the center of the waist. Across the inner arcminute, the jet is extended along the northeast-southwest direction. The direction of the jet relative to the molecular emission is shown in the third column of the channel maps in Fig.~\ref{linemap}. In Fig.~\ref{bright}, the different molecular features are overplotted on the H$\alpha$ VLT-SPHERE/ZIMPOL image of the inner jet and the two stars \citep{schmetal17}. The scenario with a relatively confined circumstellar envelope created by the recent mass loss from the AGB star, and with somewhat larger expanding arcs or rings centered on positions along the orbit (described in Sect.~\ref{wind}), seems consistent with the previously known dynamical features. In particular with that the system is known to exhibit episodic events with enhanced mass loss, for instance, the large bipolar nebula is thought to be created by a nova-like eruption $\sim$1000 years ago.

%---------------------------------------------------------------------

\section{Conclusions}
\label{conc}
We have presented the first spatially resolved observations of the thermal CO line emission from the R~Aqr system. We also reported on the line emission from other molecules and isotopologues ($^{13}$CO, $^{28}$SiO, $^{29}$SiO, and $^{32}$SO) covered by the bandwidth of the observations. The extent of the CO gas distribution is larger than previously estimated, giving a new, slightly lower estimate of the average mass-loss rate of the Mira star of about $\dot{M}$=2$\times$10$^{-7}$\,M$_{\odot}$\,yr$^{-1}$. The full extent of the CO(3-2) line emission is about 1.5-2\arcsec , and with a beam of about 0\farcs3, it would be bold to try to draw any firm conclusions about the dynamical history of molecular gas of the system. However, the spatial distributions seen at different velocities (Fig.~\ref{linemap}, Sects~\ref{dist}, and~\ref{wind}) can be reconciled with what is known from observations of the atomic gas and the orbit of the binary pair. One suggestion given is that an episode of increased mass loss ($\sim$100-50 yr ago) has created an extended ring-like structure in the orbital plane that is somewhat offset from the center component created by the recent mass loss from the AGB component. Furthermore, the central component shifts from west to east across the velocities where it is visible. The velocity shift is consistent with the gas rotating in the orbital plane. Again, higher spatial resolution observations are necessary before any firm conclusions can be drawn.

\begin{acknowledgements}
This paper makes use of the following ALMA data: ADS/JAO.ALMA\#2012.1.00524.S. ALMA is a partnership of ESO (representing its member states), NSF (USA) and NINS (Japan), together with NRC (Canada), MOST and ASIAA (Taiwan), and KASI (Republic of Korea), in cooperation with the Republic of Chile. The Joint ALMA Observatory is operated by ESO, AUI/NRAO and NAOJ. S.~M. gratefully acknowledges the receipt of research funding from the National Research Foundation (NRF) of South Africa. W.~V. was supported by ERC consolidator grant 614264. W.~V. and T.~K. further acknowledge support from the Swedish Research Council.
\end{acknowledgements}

\appendix
\section{Material moving perpendicular to the orbit}
We also report the detection of a bright (12$\sigma$) spot some distance ($\sim$7\arcsec) south of the continuum peak and peaking -22\,km\,s$^{-1}$ from the system center velocity. It is found in a direction perpendicular to the east-west elongation of the main emission component, but not precisely along the direction of the jet. There is no corresponding feature detected at positive velocities. Assuming that the orbit inclination is 70$^{\circ}$ , as derived by \citet{grommiko09} and that this is a gas clump moving perpendicular to the orbit, its deprojected velocity is $\sim$65\,km\,s$^{-1}$ \citep[almost a factor of 4 slower than the bullets ejected from V~Hya,][]{sahaetal16}. The origin of this emission component will have to be investigated by further observations.

\begin{figure}
\centering
\includegraphics[width=\columnwidth]{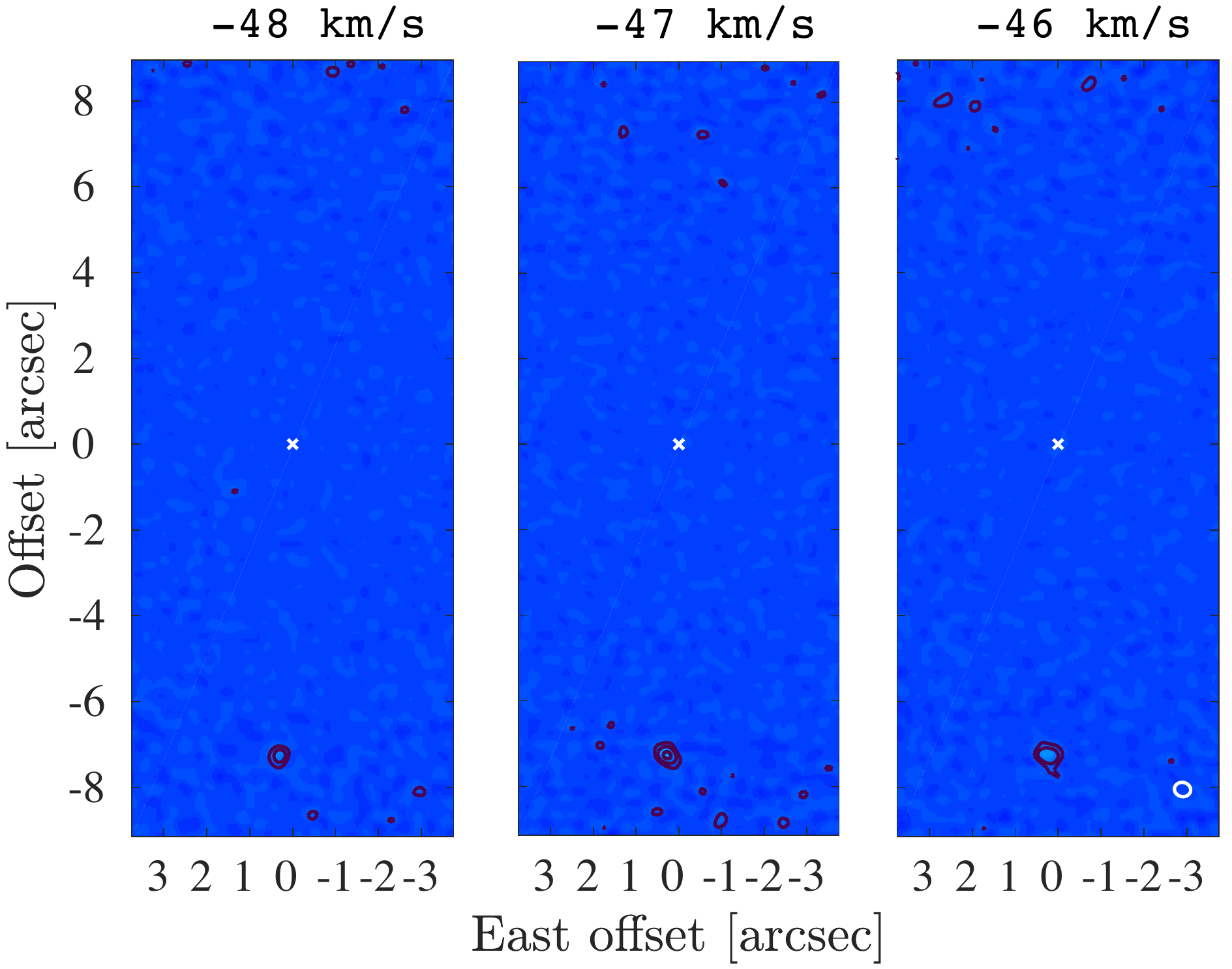}
\caption{Contours show the CO(3-2) emission spot offset $\sim$7\arcsec to the south of the system, and detected at the blueshifted line-of-sight velocities indicated in the figure. Contours are the same levels as in Fig.~\ref{linemap}. We note the spectral resolution at 1\,km\,s$^{-1}$. }
\label{spot}
\end{figure}

\bibliographystyle{aa}
\bibliography{raqr}

\end{document}